\documentclass[twocolumn]{autart} 

\usepackage{graphicx}
\usepackage{xcolor}
\usepackage{comment}
\usepackage{amsmath,amssymb,amsfonts}
\usepackage{algorithm}
\usepackage{algpseudocode}
\usepackage{soul}
\usepackage[normalem]{ulem}
\usepackage{enumerate}

\def\CD#1{\textcolor{red}{#1}}

\def\d{\mathrm{d}}
\def\N{\mathbb{N}}

\def\hxo{\widehat{x}_0}

\newtheorem{definition}{Definition}
\newtheorem{proposition}{Proposition}
\newtheorem{theorem}{Theorem}

\newtheorem{lemma}{Lemma}
\newtheorem{corollary}{Corollary}
\newtheorem{remark}{Remark}
\newtheorem{assumption}{Assumption}
\newtheorem{problem}{Problem}

\begin{document}

\begin{frontmatter}

\title{Combining Off-White and Sparse Black Models\\ in Multi-step Physics-based Systems Identification
}


\author[det,cnr]{Cesare Donati}\ead{cesare.donati@polito.it},    
\author[cnr]{Martina Mammarella}\ead{martina.mammarella@cnr.it},               
\author[cnr]{Fabrizio Dabbene}\ead{fabrizio.dabbene@cnr.it},  
\author[det]{Carlo Novara}\ead{carlo.novara@polito.it},  
\author[psu]{Constantino Lagoa}\ead{cml18@psu.edu}  

\address[det]{DET, Politecnico di Torino, Corso Duca degli Abruzzi 24, Torino, Italy}  
\address[cnr]{CNR-IEIIT, c/o Politecnico di Torino, Corso Duca degli Abruzzi 24, Torino, Italy}             
\address[psu]{EECS, The Pennsylvania State University, University Park, PA, USA}        
%

\begin{abstract} 
In this paper, we propose a unified framework for identifying interpretable nonlinear dynamical models that preserve physical properties. The proposed approach integrates physical principles with black-box basis functions to compensate for unmodeled dynamics, ensuring accuracy over long prediction horizons and computational efficiency. Additionally, we introduce penalty terms to enforce physical consistency and stability during training.
We provide a comprehensive analysis of theoretical properties related to multi-step nonlinear system identification, establishing bounds on parameter estimation errors and conditions for gradient stability and sparsity recovery. The proposed framework demonstrates significant potential for improving model accuracy and reliability in various engineering applications, making a substantial step towards the effective use of combined off-white and sparse black models in system identification. The effectiveness of the proposed approach is demonstrated on a nonlinear system identification
benchmark.
\end{abstract}

\end{frontmatter}

\section{Introduction}
\subsection{Overview and motivations}
In modern engineering applications (e.g., aerospace, automotive, energy or systems biology), when one is faced with the problem of modeling physical systems,  the classical approach is to use linear dynamical models. However, this approach has quite a few limitations.
Indeed, most of the systems that one encounters in real life exhibit dynamical behaviors which {may be} too complex to be captured by linear relationships. 
For this reason, in the last few decades, the field of nonlinear system identification has ``surged'' and many approaches have been developed with the aim of identifying nonlinear dynamical models from collected data.
However, while several important developments have been proposed, it still largely remains an open issue, as well discussed for instance by L.\ Ljung in his survey \cite{ljung2010perspectives}, and more recently in \cite{ljung2019CSM}.

\vspace{-3mm}
Existing approaches to {nonlinear systems identification} can be classified into two main groups. On one side, methods arising from \textit{basic principles}, in which the model is directly derived from the knowledge of the physical laws governing the observed system and of the relationships between the subsystems composing it. In particular, when the values of the physical parameters entering into the systems (e.g., masses or inertia of a spacecraft, or reaction coefficients of a chemical process) are themselves derived from separate {dedicated} measurements or are somehow known, one is considering a so-called \textit{white-box} model. This is however a rather extreme situation: more generally, some of the physical parameters may still need to be identified from data, and one enters in the wide family of the so-called \textit{grey-box} models. More specifically, adopting the classification proposed by Ljung, who introduced in  \cite{ljung2010perspectives,ljung2019CSM} different ``shades of grey" to distinguish the levels of knowledge of the system parameters, we refer to this particular class of models as \textit{off-white}.

On the other side of the spectrum, we have \textit{black-box} models, which aim at describing the system's dynamics using some generic parameterization or some 
family of universal approximators. Notably, this second class of models has gained increasing popularity for several important reasons. 
First, by using proper basis functions, one can easily obtain linear-in-the-parameters models which, combined with formulations aimed at minimizing the one-step prediction error, can lead to \textit{convex} problems, and allow for the derivation of several important properties of the obtained model. 
Additionally, such models can adapt to a wide range of systems with minimal (or no) prior information about the underlying physical processes. 
{Notably, }a review of the literature shows that most of the works published focus on some variation of the black model approach  (see, e.g., \cite{Chiuso-SysID-ML} and references therein) while few methodological works specifically discuss off-white approaches 
(see for instance \cite{schittkowski2002}). 
However, despite the desirable properties of linearly parameterized models, they have shown  several limitations, as for instance the lack of shared guidelines for the choice of the ``right'' basis functions and of the model order. Moreover, applications such as model predictive control have increased the need of deriving models able to provide reliable estimates over long time-intervals, calling for solutions able to provide multi-step error minimization guarantees. Clearly, such type of cost {may easily} destroy nice convexity properties, thus leading to hard optimization problems.  

To deal with these issues, the community has adopted and adapted solutions from the machine learning and artificial intelligence literature: on the one side, non-parametric approaches \cite{greblicki2008nonparametric,zorzi2017sparse}, and specially kernel-based methods \cite{chiuso-kernel} have gained popularity for their ability to capture a large diversity of nonlinear behaviors without requiring complicated choices of basis functions; on the other side, techniques based on neural networks have been introduced, showing an impressive capability of recovering long-term behaviors, and remarkable implementation ease. 
The number of works using different forms of NN-based system identification is steadily growing, and they are fast becoming the solution-to-go in several application fields. Various methods, such as Recurrent Neural Networks (RNNs), Long Short-Term Memory networks (LSTMs), Gated Recurrent Unit (GRUs), and Echo State Network (ESNs), \cite{scattoliniRNN} are being employed to capture temporal dependencies, handle long-range correlations, and spatial features, respectively, thereby enhancing the modeling of complex and dynamic systems.
Lately, however,  these techniques have also revealed their main limitations: the need of large amount of training data, and the difficulty of capturing some inherent physical phenomena at the basis of such data. The proposed solution to these two drawbacks has been the same: {to devise ways to ``bring the physics back" into the model}. This has led to the exponential growth of the family of physics-informed NNs (PINNs).
Indeed, the main feature of PINNs \cite{karniadakis2021pbnn} is exactly enabling the incorporation of physical information through either physics-based loss functions \cite{GOKHALE_PINN_loss} or structural modifications, ensuring physical consistency between inputs and outputs~\cite{JonesC_PCNN}. 

The present paper has been motivated by the fact that, although there have been significant advances in the area of system identification, there is still a need to develop identification algorithms that i) can leverage all the information available on a system such as partial parametric description, (possibly sporadic) measurements and available information on intrinsic properties of the system and ii) can make use of recent developments on first-order optimization methods, {which are at the core of the success of NN-based methods}.
With these objectives in mind, in this paper, we take a substantial step towards using off-white models in the context mentioned above, and propose a simple but efficient technique for training such models with a threefold goal: i) \textit{physical interpretability} -- we aim at deriving models whose parameters have a physical meaning, and whose values are as close as possible to the ``real" ones, ii) \textit{long prediction horizon} -- the derived model should minimize some multi-step prediction cost, to be immediately exploitable for prediction-based control schemes, and iii) \textit{computational efficiency} -- the method is based on a simple gradient scheme following the same philosophy of back-propagation-based NN training and that can be easily adapted to make use of recent results in first-order optimization algorithms.

\subsection{Setting and contributions}
\begin{figure}[!tb]
    \centering
    \includegraphics[width = 0.9\columnwidth]{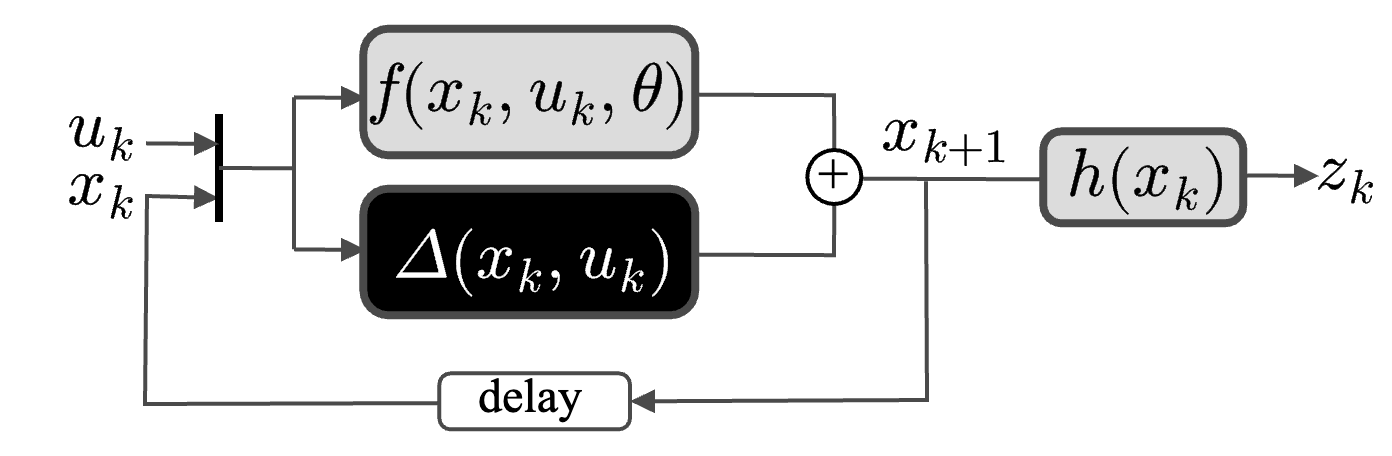}
    \caption{Dynamical system of interest.}
    \label{fig:scheme}
\end{figure}

To achieve the above goals, we introduce a unified framework that allows to identify \textit{interpretable} models of dynamical systems described by nonlinear physics-based equations 
with model uncertainties, while preserving physical properties.
The dynamical system of interest is assumed to be of the form depicted in Fig. \ref{fig:scheme}.
Here, the functions~$f$ and~$h$ are known and obtained from physical principles, while 
$\Delta$ is a discrepancy between the known physics and the dynamics resulting, e.g., from modeling errors, aging, uncertainties, and perturbations existing in any realistic problem. 

Despite the knowledge of physical laws, the coherence between physics and estimated parameters may not be ensured.
Indeed, they might not fully capture specific system behaviors, such as passivity, monotonicity, divergence, symmetry, and stability, that must be considered to guarantee adherence to physical principles \cite{ferraritrecate2022physicalconstraints}.
In such situations, the proposed framework introduces specific penalty terms in the cost function to reflect the physical characteristics of the system's behavior.
Additionally, some systems may exhibit finite escape time phenomena for certain parameters or initial condition values, leading to an explosive behavior and consequent exploding gradients. This motivates the use of penalties such as barrier functions on the states and parameters, to avoid this issue during the training and force the estimated parameters to generate trajectories that are bounded in finite time. This allows the training process to remain in a ``safe" neighborhood around the measured trajectory. 

Moreover, we make use of a mixed approach that combines the multi-step identification method introduced in \cite{donati2023oneshot} with a sparse black-box component designated to compensate for unmodeled dynamics in the physical model. By integrating partial knowledge of the nonlinear system's physical description with black-box basis functions, the resulting structure effectively counterbalances the limitations of the available dynamical model. The black-box component serves as a robustifier capturing unforeseen dynamics and model oversimplification, enhancing overall accuracy and reliability, not only from an output estimation error point of view but also it often leads to better estimates of the parameters of the parametric part of the model. Indeed, if one does not use a black box component, the presence of unmodeled  
components often affects the accuracy of the physical parameters. This happens when the identification algorithm tries to encompass the residual dynamical effects caused by the unmodelled part in the identified physical parameter value, with the aim of minimizing the prediction error. This comes at the cost of unreliable identified physical values \cite{mammarella2024blended}. Moreover, the interest in finding the sparsest coefficient vector selecting the basis functions is driven by the necessity of having a compensation term that only deals with the unmodeled dynamics and it is easily ``interpretable''. Indeed, on the other hand, non-sparse coefficient vectors might capture part of the dynamics of the known physical model, resulting in predictions that align well with the measurements, but with physical parameters that deviate from the true nominal values.
A similar model augmentation strategy was presented in \cite{liu2024physics}, where prior physics-based state-space models are augmented using a weighted function regularization. However, differently from our work, the physical parameters are assumed to be known and are not part of the identification. In contrast, our focus is on accurately identifying the physical parameters while leveraging black-box augmentation models. 

Lastly, exploiting the structural definition of the proposed framework, we briefly introduce a linear time-varying dynamical system that models the gradient evolution along the prediction horizon. This is achieved by exploiting the recursive and physics-based structure of the proposed framework, which allows for a structured decomposition of the problem into subproblems that can be solved using techniques typically used for Neural Networks such as Automatic Differentiation (AD) \cite{baydin2018autodiffsurvey}. This approach enables us to characterize the gradient evolution for our specific physics-based structure intrinsically linking it to the system we are identifying.

To the best of our knowledge, within this general framework, we provide for the first time a comprehensive analysis of several crucial theoretical properties regarding multi-step, nonlinear system identification. Specifically, we establish theoretical bounds on the parameter estimation error and sufficient conditions for the stability of the gradient, ensuring that the optimization process remains consistent and reliable. Furthermore, we analyze the conditions necessary for the exact recovery of the sparsity of the black-box term, which is essential for accurately capturing the underlying structure of the considered system. 

\subsection{Paper organization}
The remainder of the paper is structured as follows. In Section~\ref{sec:framework}, we formulate the problem and define the identification framework, introducing the main features of the considered system dynamics and of the estimation model. In particular, Section~\ref{sec:bbcomp} details how unmodeled dynamics can be compensated with a regulated black-box model, while Section~\ref{sec:pbconstraints} describes the approach used to enforce possible physics-based constraints based on prior knowledge of the system. The dynamical system describing the evolution of the gradient is detailed in Section~\ref{sec:RGC}. A theoretical analysis of the proposed framework is presented in Section~\ref{sec:theo}, while Section~\ref{sec:sparsityrecovery} shows an analysis on the sparsity of the black-box term.
Numerical results on two examples are discussed in Section~\ref{sec:num_res}, while main conclusions are drawn in Section~\ref{sec:concl}.

\subsection{Notation} 
{The $\ell_p$ norm of a vector $v$ is denoted as $\|v\|_p$, with $\|v\|$  the Euclidean norm. The spectral norm of a matrix $A$ is denoted  as $\|A\|$.
Given a vector $v$,
we denote by $\mathbf{v}_{1:T}\doteq\{{v_k}\}_{k=1}^{T}$  the sequence of vectors $\{v_1,\ldots,v_T\}$, and $\|\mathbf{v}_{1:T}\|_p \doteq \|[v_1^\top,\ldots,v_T^\top]^\top\|_p$.
Given $\mathbf{v}_{1:T}$ and a function $f(v_k)$ we define $f(\mathbf{v}_{1:T}) \doteq \left[f(v_1),\dots, f(v_T)\right]^\top$ the vector composed by the evaluations of the function in every element of the sequence.}
Given integers $a,b\in\N, a\leq b$, we denote by $[a,b]$ the set of integers $\{a,\ldots,b\}$.  The vectorization of a matrix $A\in \mathbb R^{m,n}$, denoted $\text{vec}(A)$, is the $mn \times 1$ column vector obtained by stacking the columns of the matrix $A$ on top of one another. The support of a vector is defined as the set of indices at which it is not null, i.e., $\mathrm{supp}(v)~\doteq~\{i~:~v_i\neq0\}$. Similarly, we define $\overline{\mathrm{supp}}(v)$ as the set of indices at which it is null, i.e., $\overline{\mathrm{supp}}(v)~\doteq~\{i~:~v_i=0\}$.
\section{{System identification framework}} \label{sec:framework}
Let us consider a dynamical system $\mathcal{S}$, described as shown in Figure \ref{fig:scheme} by a combination of a physical model and an unknown model, with the following state equations
\begin{equation}
    \begin{aligned}
    \mathcal{S}:\quad&x_{k+1} = f\left({x}_k, {u}_k; \theta\right) + \Delta(x_k,u_k),\\
    &z_k = h\left(x_k\right).
    \end{aligned}
\label{eqn:system}
\end{equation}
The system is described by a state vector $x \in \mathbb{R}^{n_x}$, and a vector of parameters $\theta \in \mathbb{R}^{n_\theta}$, and it receives as input the vector $u \in \mathbb{R}^{n_u}$, while $z \in \mathbb{R}^{n_z}$ is the vector of observations. 
The unknown system parameters, i.e., $\theta$, and the unknown initial conditions, i.e., $x_0$, must be estimated.
The known functions $f:\mathbb{R}^{n_x}\times\mathbb{R}^{n_u}\times \mathbb{R}^{n_{\theta}}\to\mathbb{R}^{n_{x}}$, and~$h:\mathbb{R}^{n_x}\to\mathbb{R}^{n_{z}}$, represent part of the state update function and the observation function, respectively. They are assumed to be nonlinear, time-invariant, and continuously differentiable\footnote{{The formulation  can be  extended to include time-varying systems.}}. On the other hand,~${\Delta:\mathbb{R}^{n_x}\times\mathbb{R}^{n_u}\to\mathbb{R}^{n_{x}}}$ is not known and {will} be estimated using a black-box approximation. 
To identify $\theta$, $x_0$, and $\Delta$, sequences of collected inputs $\mathbf{{\widetilde u}}_{0:T}$ and corresponding observations $\mathbf{{\widetilde z}}_{0:T}$ are used, where for each $k\in[0,T]$ we have
\begin{equation}
\widetilde u_k = u_k + \eta^u_k,\quad \widetilde z_k = z_k + \eta^z_k,
\label{eqn:measurements}
\end{equation}
with $\eta^u_k$ and $\eta^z_k$ being the input and output measurement noises, respectively. 

The goal is to identify $\theta$ and $x_0$, thus defining an estimation model $\mathcal{M}$ approximating the system $\mathcal{S}$, while compensating for the unmodeled dynamics if $\Delta \neq 0$.
While in a standard single-step framework the model is applied once to each input to predict the states one step ahead in time, a multi-step identification problem involves the propagation of the prediction of ${x_k}$ over the desired horizon $T$, recursively applying the dynamical model $\mathcal{M}$ \cite{forgione2023from}.
Specifically, by introducing a cost function $\mathcal{C}_T$ measuring the error along the prediction horizon (and, possibly, additional penalty terms to take into account, e.g., physically-driven or sparsity constraints), we state the following multi-step identification problem.
%
\begin{problem}[multi-step identification]\label{prob:msidproblem}
    Given the state and output functions $f$ and $h$ in \eqref{eqn:system}, a sequence of~$T$ collected inputs $\mathbf{{\widetilde u}}_{0:T}$ applied to $\mathcal{S}$, {along with} the corresponding~$T$ collected observations $\mathbf{{\widetilde z}}_{0:T}$, identify the optimal values for the physical parameters $\theta$ and initial condition $x_0$ over the simulation horizon $T$, such that~$\mathcal{M}$ is the physically-consistent best approximation of~$\mathcal{S}$. In particular, given a multi-step cost function~$\mathcal{C}_T$, we aim at solving the following optimization problem\footnote{{To guarantee that the minimizer is unique, some (local) identifiability assumptions are needed, see Section~\ref{sec:identifyability}.}}
    \begin{equation}
        \left( {\theta}^\star, x_0^\star\right) \doteq \arg \min_{\theta, x_0}\,\, \mathcal{C}_{T}.
        \label{eqn:optprobl}
    \end{equation}
\end{problem}

In the following sections, we design the multi-step cost function $\mathcal{C}_T$ according to specific choices of the estimation model $\mathcal{M}$.
Specifically, in Section \eqref{sec-OW}, we address the simpler scenario where $\Delta = 0$ and propose a potential choice for $\mathcal{M}$ and $\mathcal{C}_T$. Then, in Section \eqref{sec:bbcomp}, we extend the discussion to the case of $\Delta \neq 0$. {Furthermore}, we will present how the cost function can be additionally tailored to enforce physics-based behavior.
\subsection{Multi-step off-white model}
\label{sec-OW}
\begin{figure}[!ht]
    \centering
    \includegraphics[width = 0.9\columnwidth]{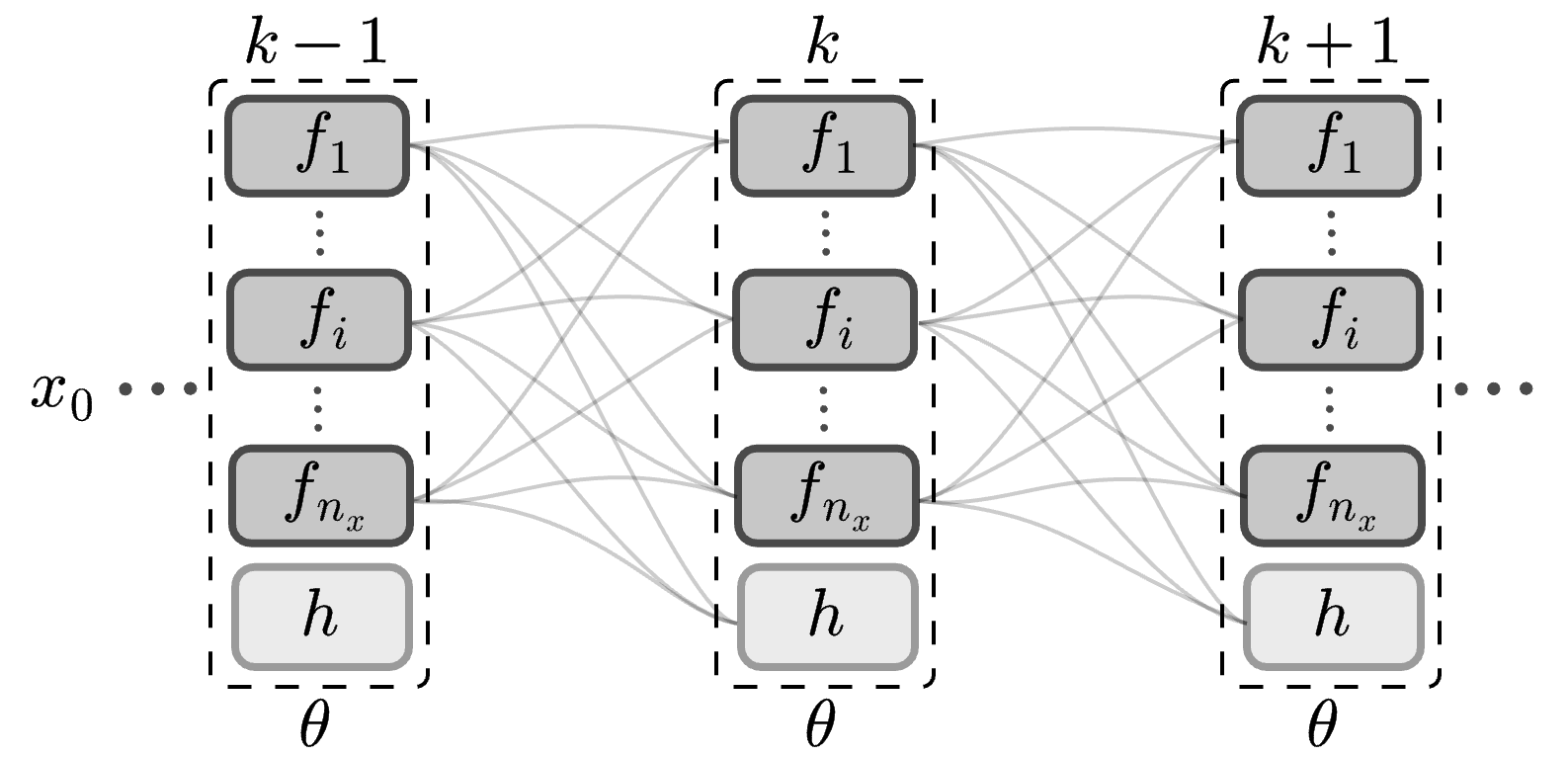}
    \caption{Multi-step, off-white model structure.}
    \label{fig:NNlike}
\end{figure}

We start considering the framework with $\Delta = 0$, which leads to a fully-known system and to a straightforward choice for the off-white 
 dynamical model $\mathcal{M}$
\begin{equation}
\begin{aligned}
    {\mathcal{M}}\,:\quad\widehat{ x}_{k+1} &=  f(\widehat{ x}_{k},  {u}_{k}; \widehat \theta),\\
    \widehat {{z}}_{k} &= {h}(\widehat{x}_{k}),
    \label{eqn:model.simple}
    \end{aligned}
\end{equation}
where $\widehat{x}_{k} \in \mathbb{R}^{n_x}$, $\widehat{z}_{k} \in \mathbb{R}^{n_z}$ are the estimated state and observation at time $k$ obtained using the identified vector of parameters $\widehat \theta \in \mathbb{R}^{n_\theta}$ and initial condition $\hxo \in \mathbb{R}^{n_x}$. 
The multi-step propagation of the model in \eqref{eqn:model.simple} is depicted in Fig.~\ref{fig:NNlike}.

In this case, the multi-step cost function~$\mathcal{C}_{T}$ can be defined as the sum of $T$ local losses $\mathcal{L}_k\left(\theta,x_0;e_k\right)$, i.e.,
\begin{equation}
\mathcal{C}_{T}\left(\theta,x_0;\mathbf{e}_{0:T} \right) = \sum_{k=0}^{T} \mathcal{L}_k\left(\theta,x_0;e_k\right).   
    \label{eqn:cost}
\end{equation}
Note that the cost in \eqref{eqn:cost} is a function of $\theta$ and $x_0$, and depends on the prediction error sequence 
\begin{equation}
\mathbf{e}_{0:T}=\{e_k\}_{k=0}^T \text{ with } e_k\doteq\widehat z_k-{\widetilde z_k}\in\mathbb{R}^{n_z},\, k \in [0,T].
    \label{eqn:err}
\end{equation}
More precisely, the cost depends on the initial state $x_0$ and on the collected measurements through
\eqref{eqn:model.simple}, which, in turn depends on the noise sequence  $\boldsymbol{\eta}=\{\boldsymbol{\eta}_{0:T}^{u},\boldsymbol{\eta}_{0:T}^{z}\}$, with~$\boldsymbol{\eta}_{0:T}^{u}=\{\eta_{k}^{u}\}_{k=0}^{T}$,~$\boldsymbol{\eta}_{0:T}^{z}=\{\eta_{k}^{z}\}_{k=0}^{T}$.
{In the following}, depending on the context, we will highlight the considered dependencies of $\mathcal{C}_T$ by observing that
\begin{align}
\nonumber 
\mathcal{C}_{T}\left(\theta,x_0;\mathbf{e}_{0:T} \right)&=
\mathcal{C}_{T}(\theta,x_0;\mathbf{{\widetilde u}}_{0:T},\mathbf{{\widetilde z}}_{0:T})\\
&=
\mathcal{C}_{T}(\theta,x_0;\boldsymbol{\eta}).
\label{eq:CT_eta} 
\end{align}
In this paper, we assume that $\mathcal{L}_k: \mathbb{R}^{n_z}\!\times\!\mathbb{R}^{n_{\theta}} \rightarrow \mathbb{R}$ is a twice differentiable with continuous derivatives function\footnote{To simplify the notation when needed, we will drop the arguments from the local loss function $\mathcal L_k$.}.  
Notice that, the loss term in \eqref{eqn:cost} is very general, and allows for a flexible customization of the cost function in order to be adapted to different scenarios. In the upcoming sections, we show how it can be modified to properly compensate for unmodelled terms in the dynamics and incorporate physical properties.

\subsection{Black model augmentation}\label{sec:bbcomp}

\begin{figure}[!ht]
    \centering
    \includegraphics[width=0.9\columnwidth]{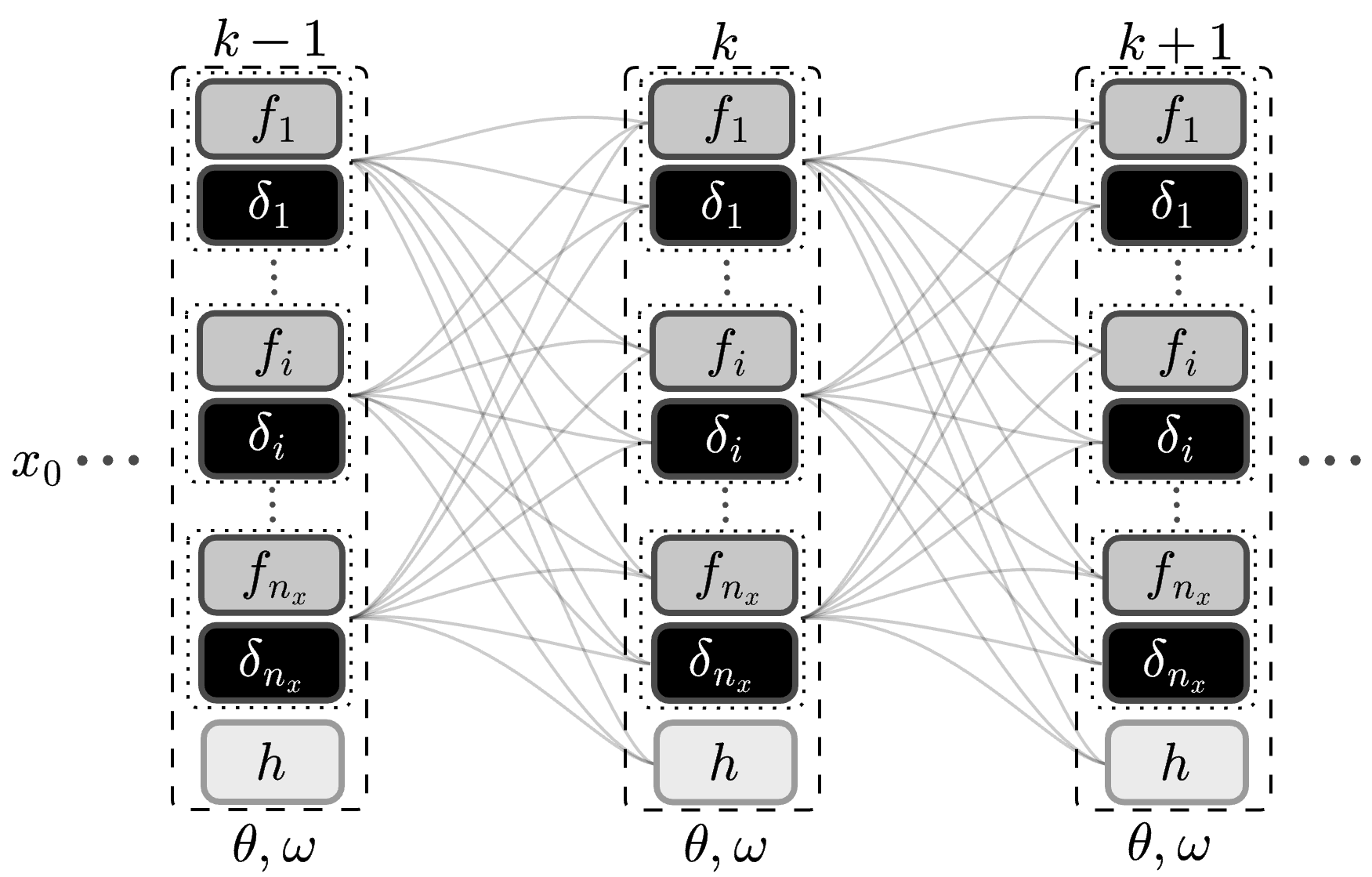}
    \caption{Multi-step combination of off-white and black models.}
    \label{fig:bb}
\end{figure}

Let us now consider the case for $\Delta \neq 0$. 
In this scenario, the estimation model is enriched by introducing a black-box component to capture the dynamics that remain unaccounted by the physical priors, as shown in Fig.~\ref{fig:bb}.

In this case, the function $\Delta$ in \eqref{eqn:system} represents an unknown part of the state equations in the system $\mathcal{S}$. In the following, we will denote with $\delta$ the \textit{model}\footnote{Note that, since no information exists on the functional form of the mathematical model defining $\Delta$, the statement~${\delta = \Delta}$ is in general not true.
} exploited to approximate the {unknown term} $\Delta$. 
The aim of the black-box term $\delta$ is to augment the incomplete physical dynamics $f$, by approximating $\Delta(x_k,u_k)$, i.e., the discrepancy between the state equation of system \eqref{eqn:system} and the available physical model. Hence, it must have only a complementary role, so that the more the available physical knowledge describes the physics of the system, the more $\delta$ will be ``small". Moreover, besides minimizing the effect of the black-box part, we also aim to explain the model discrepancies with the simplest possible model.
For this reasons, a regularization term is introduced in the cost function,
to ensure that the black-box component remains minimal when physical knowledge adequately describes the system, thus
helping to maintain the interpretability and simplicity of the model.

Thus, the proposed multi-step estimation model is defined as follows
\begin{equation}
\begin{aligned}
    {\mathcal{M}}\,:\quad\widehat{x}_{k+1} &= {f}(\widehat{x}_{k},  {u}_{k}; \widehat \theta) + {\delta}(\widehat{x}_{k},u_k; \omega),\\
    \widehat {{z}}_{k} &= {h}(\widehat{x}_{k}),
    \label{eqn:bb_extension_v2}
    \end{aligned}
\end{equation}
where 
$\delta(\widehat{x}_{k}, u_k; \omega) = [\delta_1(\cdot), \dots, \delta_{n_x}(\cdot)]^\top$, and~$\omega$ collects all the design parameters, 
to be learned during the training procedure, describing the black model (e.g., the collection of different weights defining a neural network or simply the coefficients of an ARX representation). 
%

In our setup, for simplicity of treatment, we consider a black model consisting of a combination of basis functions, but we remark that the approach may be extended to more involved families of approximators, such as neural networks \cite{forgione2021dynonet} or kernel-based methods \cite{chiuso-kernel}.
Formally, let us define $\varphi~\in~\mathbb{R}^{m}$, i.e., the vector of basis functions $\varphi(\widehat x_k, u_k)~=~\left[\varphi_1(\widehat x_k,u_k), \dots, \varphi_m(\widehat x_k,u_k) \right]^\top$, with $\varphi_j: \mathbb R^{n_x}\times\mathbb R^{n_u}\to\mathbb R$.
{The $\iota$-th element of $\delta$ is defined as}
\begin{equation}
    \delta_\iota
    = \sum^m_{j=1} \Omega_{\iota j}\varphi_j\!\left(\sum_{i=1}^{n_x}W_{ij}^{(\iota)}\hat x_{i,k}\!+\!\sum_{i=1}^{n_u}W_{(n_x+i)j}^{(\iota)}u_{i,k}\!+\!B_{\iota j}\right)\!,
    \label{eqn:bbmodel}
\end{equation}
with $\iota \in \,\left[1,n_x\right]$. Then, we have $\omega = \left\{\Omega,W,B\right\}$,
with $\Omega=[\Omega_{\iota j}]~\in~\mathbb R^{n_x,m}$ collecting the weights of the basis functions for every state, and
$W = [W^{(1)},\dots, W^{(n_x)}] \in \mathbb R^{(n_x+n_u),n_xm}$ with $W^{(\iota)} = [W_{ij}^{(\iota)}] \in \mathbb R^{n_x+n_u,m}$, and $B =[B_{\iota j}]\in \mathbb R^{n_x,m}$ collecting additional weighting terms, i.e., the coefficients of the states and inputs, and the bias appearing in the argument of each basis functions.

{As discussed}, we aim at minimizing the number of different basis adopted to describe it, thus aiming at sparsifying the term $\Omega$. It is well-known 
\cite{calafiore2020sparse} that a classical regularization term that can be introduced to sparsify the black-box weights is the $\ell_1$ norm, which is chosen in this paper for its ability to promote sparsity and ensure a simpler and more interpretable representation of the unmodeled dynamics. However, it should be noted that other norms can also be selected, depending on the specific requirements and characteristics of the application.
Let us consider the modified {multi-step regression cost} $\mathcal{C}_T$ as the sum of local losses over the prediction horizon $T$ accounting for the black-box term regularization, i.e.,
\begin{equation}
    \mathcal{C}_T(\theta,x_0,\omega; \mathbf{e}_{0:T}) \doteq
    \sum_{k=0}^{T} \mathcal{L}_k + \gamma \sum_{\iota=1}^{n_x} \|\Omega_\iota^\top\|_1,
    \label{eqn:cost_general}
\end{equation}
with $\Omega_\iota^\top$ the $\iota$-th row of $\Omega$, and $\gamma \in \mathbb R$ penalizing the regularization term such that the lower the sparsification of~$\Omega$ rows is, the larger the associated black-box loss value will be.
Notice that, since the $\ell_1$-norm is not differentiable, it will be replaced by a differentiable approximation of the absolute value to allow gradient computation. In this work, we exploit the softplus approximation of the absolute value \cite{schmidt2007fast}
such that, for a vector~${v \in \mathbb R^{n_v}}$, the approximated $\ell_1$-norm is
$$
\|v\|_1 \approx {\|v\|_{1\beta}}\doteq\frac{1}{\beta}\sum_{i=1}^{n_v} [\log (1+ \mathrm{e}^{-\beta v_i})+\log (1+ \mathrm{e}^{\beta v_i})],
$$
where the tunable parameter $\beta$ controls the level of approximation. Thus, we define the following cost function,
\begin{equation}
\mathcal{C}_T(\theta,x_0,\omega; \mathbf{e}_{0:T}) \doteq
    \sum_{k=0}^{T}\left(\mathcal{L}_k + {\gamma}\|\Omega\|_{1\beta}\right).
    \label{eqn:reg_term}
\end{equation}
\begin{remark}[subgradient approach]
Instead of using an $\ell_1$~-~norm approximation, it is also possible to exploit a subgradient approach {\cite{nesterov2009primal}} to deal with the gradient of a non-differentiable cost function. This is left out in this paper for space reasons.
\end{remark}
Therefore, we recast Problem~\ref{prob:msidproblem} as follows.
\begin{problem}[augmented multi-step identification]
\label{prob:msidproblem.BB}
Given $f$ and $h$ in \eqref{eqn:system}, sequences $\mathbf{{\widetilde u}}_{0:T}$ and $\mathbf{{\widetilde z}}_{0:T}$, and the vector of basis functions $\varphi$, estimate the optimal values for the physical parameters $\widehat \theta$, initial conditions $\widehat x_0$, and the black-box weights $\omega$ over the simulation horizon $T$, such that the model $\mathcal{M}$ of the form defined in \eqref{eqn:bb_extension_v2} is the physically-consistent best approximation of $\mathcal{S}$. Thus, given the multi-step cost function $\mathcal{C}_T$ we aim at solving the following optimization problem
\begin{equation}({\theta}^\star,  {x}_0^\star, \omega^\star) \doteq \arg \min_{\theta, x_0, \omega}\,\, \mathcal{C}_{T}.
    \label{eqn:optprob.bb}
\end{equation}
\end{problem}

%

%
\subsection{Physics-based penalty functions}\label{sec:pbconstraints}
One of the objectives of this work is to incorporate physical properties to the identified model $\mathcal{M}$, either \eqref{eqn:model.simple} or \eqref{eqn:bb_extension_v2}. To this end, besides explicitly using the physical laws in defining the model equations, we may also follow the same philosophy adopted in many PINNs based approaches \cite{karniadakis2021pbnn,GOKHALE_PINN_loss}, i.e.,  to integrate physical constraints directly into the cost function, enabling a ``\textit{physics-guided learning}" that also leverages domain knowledge to inform the optimization process. In this way, the optimization is ``steered" towards physically-consistent and meaningful solutions, thereby enhancing the robustness and reliability of the optimization outcome.

Specifically, this is achieved by inducing physical constraints through additional \textit{penalty terms} introduced directly into the cost function $\mathcal{C}_T$, in order to keep the optimization problem unconstrained, as
\begin{equation}
\mathcal{C}_T(\theta, x_0; \mathbf{e}_{0:T},\widehat{\mathbf{x}}_{1:T})\!=\!\sum_{k=0}^{T} \left(\mathcal{L}_k\!+\!\lambda p(\widehat x_k, \theta)\!+\!\nu q^2(\widehat x_k, \theta)\right),
\label{eqn:cost_general_phy}
\end{equation}
for the case of $\Delta = 0$, and 
\begin{equation}
    \begin{aligned}
&\mathcal{C}_T(\theta,x_0,\omega; \mathbf{e}_{0:T},\widehat{\mathbf{x}}_{1:T})\\
 & \qquad\doteq
    \sum_{k=0}^{T} \left( \mathcal{L}_k + \lambda p(\widehat x_k, \theta)+ \nu q^2(\widehat x_k, \theta) + {\gamma}\|\Omega\|_{1\beta} \right),
    \end{aligned}
    \label{eqn:final_cost}
\end{equation}
for the case of $\Delta \neq 0$. Here, $\lambda \in \mathbb R$ and $\nu \in \mathbb R$ are the penalty weights, while $p, q: \mathbb{R}^{n_x} \times  \mathbb{R}^{n_\theta} \rightarrow \mathbb{R}$ are time-invariant, continuously twice differentiable functions such that 
$p(\widehat x_k, \theta) \leq 0$, $q(\widehat x_k, \theta) = 0$ $\forall k \in [0, T]$.
By customizing the functions $p$ and $q$, a range of physics-based constraints can be induced in the optimization problem. In the following, two case studies are presented. More details and examples such as energy conservation, symmetry of the inertia matrix, positive semi-definiteness, and the triangle inequality can be found in \cite{donati2023oneshot} and \cite{mammarella2024blended}. 

\subsubsection{Convex constraints set in the parameters space}
The first class of constraint we consider is rather natural, and consists in capturing
physical limits on the model parameters $\theta$. That is, we assume {there exists  upper bound $\theta^{ub} = [\theta^{ub}_i]$,
and lower bound $\theta^{lb} =[\theta^{lb}_i]$,
$i\in [1,n_\theta]$, with $\theta^{ub}_i,\theta^{lb}_i$ finite,} such that 
\begin{equation}
    \theta \in \Theta \doteq \left\{\theta^{lb}_i \leq \widehat \theta_i \leq \theta^{ub}_i,\,\, i = [1,n_\theta]\right\}.
    \label{eqn:cvx_set}
\end{equation}
An \textit{exponential barrier function} can be used to define the physics-based penalty term as follows
\begin{equation}
    p(\widehat{x}_k, \theta) \doteq \|e^{\alpha(\hat{\theta} - \theta^{ub})}\|_2^2 + \|e^{\alpha(\theta^{lb} - \hat{\theta})}\|_2^2,
    \label{eqn:expbarriertheta}
\end{equation}
where the positive scalar $\alpha \in \mathbb R$ represents a sharpness parameter, encouraging estimated parameter variables to remain within known intervals\footnote{{Alternatively, whenever $\Theta$ is a convex set, the identification algorithm can be modified by incorporating a projection step immediately after the parameters update following the gradient computation. We do not pursue this solution since it could in principle hinder first-order algorithms convergence.}}.
Here, a special case is the {parameter non-negativity constraint}, where~${\widehat \theta_{i,k} \geq 0}$ for all $k$, and \eqref{eqn:expbarriertheta} becomes
~$p(\widehat{x}_k, \theta) \doteq \|e^{-\alpha\hat{\theta}_k}\|_2^2.$

%
\subsubsection{Physical limits}
In most cases, specific bounds for parameter values may not be known. However, physical limits of the state variable can be used to ensure that the model is physically-consistent. Thus,
similar bounding constraints are defined to enforce limits on the state variable being predicted. For instance, the constraint
\begin{equation}
    x \in \mathcal{X} \doteq \left\{x^{lb}_i \leq \widehat x_i \leq x^{ub}_i,\,\, i = [1,n_x]\right\},
    \label{eqn:cvx_set_x}
\end{equation}
can be expressed with the following penalty term
$$p(\widehat{x}_k, \theta) \doteq \|e^{\alpha(\hat{x}_k - x^{ub})}\|_2^2 + \|e^{\alpha(x^{lb} - \hat{x}_k)}\|_2^2.$$
encouraging the predicted state variables to stay within defined physical limits, or specific intervals where the trajectories are known to be stable. This aspect will be  further {discussed} in Section \ref{sec:explgrad}.

\section{Direct gradient computation and gradient dynamics} \label{sec:RGC}
As discussed in the introduction, {an important contribution of this work is also to provide \textit{computable tools}} for 
exploiting first-order methods for the solution of the optimization problem \eqref{eqn:optprobl} or \eqref{eqn:optprob.bb},  at the core of our multi-step identification procedure. Such methods, which under suitable conditions are guaranteed to converge to a solution (in general sub-optimal), have recently re-gained popularity for their ability to tackle large-scale problems. 
%
At the core of first-order techniques is the computation of the gradient of the cost function $\mathcal{C}_T$ evaluated at the current solution. For instance, in
 standard gradient descent, once the gradients
with respect to the parameter $\nabla_{\theta}\mathcal{C}_{T} \in \mathbb R^{n_\theta}$ and the initial condition $\nabla_{x_0} \mathcal{C}_{T}  \in \mathbb R^{n_x}$ 
are computed, the weights are updated in the direction that minimizes the total cost, i.e.,
\begin{subequations}
\begin{align}
    \widehat \theta &\gets \widehat \theta - \varsigma_{\theta} \nabla_{\theta} \mathcal{C}_{T},
    \label{eqn:wupdate}\\
    \hxo &\gets \hxo - \varsigma_{x_0} \nabla_{x_0} \mathcal{C}_{T},
    \label{eqn:xoupdate}
\end{align}
\label{eqn:GD}
\end{subequations}
where $\varsigma_{\theta}$ and $\varsigma_{x_0}$ are  scalar learning rates, that can be designed exploiting  state-of-the-art methods {\cite{behera2006adaptive}}.

Clearly, various methods are available in the literature for calculating the gradient, with varying degrees of approximation, and, at least in principle, all of these are applicable to the proposed framework, varying from numerical differentiation to automatic differentiation \cite{baydin2018autodiffsurvey}.
%
As a first important contribution of the paper, we exploit the explicit knowledge of the function $f(\cdot)$ in \eqref{eqn:system}, and propose a novel and efficient method to compute the exact gradient. The method is inspired by the classical backpropagation scheme, adopted, e.g., in neural networks, and recently proposed in the context of system identification in our recent work \cite{donati2023oneshot}, and similarly in \cite{FerrariTrecate2023simba}. 
{However,} differently from backpropagation through time, where the error is back-propagated in time in an unfolded RNN only after forward-propagating the inputs through the unfolded network, we represent the evolution of the gradient by studying the predictions and errors of the system as they evolve forward in time. Thus, we propagate the gradient through the recursion intrinsically defined in dynamical systems updating it while propagating the predicted states. This involves the definition of a "memory" of the effects of past errors and an "innovation" due to current errors.

The solution we propose, still based on the chain rule differentiation, presents several distinguishing characteristics that make it of particular interest. First, the method still produces exact gradient values but offers a significantly improved computational performance compared to  \cite{donati2023oneshot}. This is obtained by {propagating forward in time}  the impact of $\widehat \theta$ and $\hxo$ on current predictions, exploiting the definition of a memory matrix, and thus avoiding the necessity of multiplying similar matrices as in the backpropagation scheme. 
Second, it allows the definition of a particular time-varying discrete-time dynamical system describing the evolution of the gradient with respect to parameters and initial condition. The closed-form formula for the evolution of the gradient not only reduces the computational cost but also allows to study analytically the behavior of the gradient during the identification process. Specifically, we will show in Section \ref{sec:explgrad} how it can be used to derive sufficient conditions for avoiding the phenomenon of so-called ``exploding gradient". 

Before presenting our first result, we introduce some {additional} simplifying notation: 
the Jacobian matrices of $e_k$ with respect to $z_k$ and of $z_k$ with respect to $x_k$ are denoted as~
$\mathcal{J}^{e\!/\!z}_k\doteq \frac{\partial e_k}{\partial z_k}\in\mathbb R^{n_z, n_z}$, 
and $\mathcal{J}^{z\!/\!x}_k\doteq\frac{\partial z_k}{\partial x_k}\in\mathbb R^{n_z, n_x}$,  respectively.
Similarly, $\mathcal{J}^{x\!/\!\theta}_k\doteq\frac{\partial x_k}{\partial \theta} \in \mathbb R^{n_x, n_\theta}$ is the Jacobian matrix of $x_k$ with respect to $\theta$, while $\mathcal{J}^{x\!/\!x}_k \doteq \frac{\partial x_k}{\partial x_{k-1}} \in \mathbb R^{n_x, n_x}$ is the Jacobian matrix of $x_k$ with respect to $x_{k-1}$.
Notice that $\mathcal{J}^{e\!/\!z}_k$, $\mathcal{J}^{z\!/\!x}_k$, $\mathcal{J}^{x\!/\!x}_k$, $\mathcal{J}^{x\!/\!\theta}_k$ are time-varying matrices with fixed structure, depending on the values of~$\widehat x_k, \widetilde u_k, \widehat z_k, \widetilde z_k, e_k$, $\widehat \theta$, and $\hxo$ at time-instant $k$.
For instance, we have that
\begin{equation*}
    \mathcal{J}^{x\!/\!x}_k = \frac{\partial x_k}{\partial x_{k-1}} = \frac{\partial f(\widehat x_{k-1}, \widetilde u_{k-1}, \widehat \theta)}{\partial \widehat x_{k-1}}.
\end{equation*}
The following results are first presented considering a cost function of the form \eqref{eqn:cost}. Then their extension to the final cost \eqref{eqn:final_cost} is formalized in the subsequent remark.
In the following proposition, whose proof is reported in Appendix \ref{app:prop1}, we show how to represent the evolution of the gradient with respect to $\theta$ as a dynamical system.
\begin{proposition}[gradient dynamics -- $\theta$]
\label{prop1}
{Define the memory matrix}
\begin{equation}
\label{eqn:Lambdadef}
\Lambda_{k} \doteq \frac{\d {x_k}}{\d {\theta}} \in \mathbb R^{n_x,n_\theta},
\end{equation}
as the matrix containing the total derivatives of the states with respect to $\theta$.
Define  vectors $\rho_k \in \mathbb R^{n_x}$, $\varrho_k \in \mathbb R^{n_\theta}$ as 
\begin{equation}
    \label{eqn:rho-var-def}
    \rho_k \doteq \left({\nabla_{e}^\top  \mathcal{L}_k}{\mathcal{J}^{e\!/\!z}_k}{\mathcal{J}^{z\!/\!x}_k}\right)^\top,\quad
    \varrho_k \doteq {\nabla_{\theta} \mathcal{L}_k}.
\end{equation}
The gradient evolution with respect to the parameter $\theta$ along the multi-step horizon $T$ is described by the following time-varying dynamical system
\begin{subequations}
    \begin{align}
     \label{eqn:GUlaw_mat}
     &\Lambda_{k} = \mathcal{J}^{x\!/\!x}_k\Lambda_{k-1}+
     \mathcal{J}^{x\!/\!\theta}_k, \\
     \label{eqn:GUlaw_grad}
     &\nabla_{{\theta}} \mathcal C_{k} = \nabla_{{\theta}} \mathcal C_{k-1} + \Lambda^\top_k\rho_k + \varrho_k,    
    \end{align}
    \label{eqn:GULAW}
\end{subequations}
with
    $\Lambda_{0} \doteq \frac{\d x_0}{\d \theta} = \mathbf{0}_{n_x,n_\theta}$, $\nabla_{{\theta}} \mathcal C_{0} = \mathbf{0}_{n_\theta}$.
\end{proposition}
In \eqref{eqn:GUlaw_mat} it is possible to notice the term $\mathcal{J}_k^{x\!/\!\theta}$, reflecting the impact of the parameter estimation $\widehat \theta$ on the current state prediction $\widehat x_k$ and, consequently, on $\mathcal{L}_k$. Specifically, it characterizes the ``direct" effect of $\widehat \theta$ on~$\widehat{x}_k$. Conversely, the term $\Lambda_{k-1}$ serves as memory, encapsulating how the effect of $\widehat \theta$ on past predictions, i.e.,~$\widehat x_\tau$ with $\tau\in[1,k-1]$, has affected the current state estimation $\widehat x_k$.
Thus, \eqref{eqn:GUlaw_grad} defines a formula in which the gradient is updated at each time step $k$ with an \textit{innovation term} exploiting the information encapsulated in $\Lambda_{k}$, i.e., the effect of the parameter estimation $\widehat \theta$ on the prediction horizon up to time $k$.

In the following proposition, whose proof is reported in Appendix \ref{sec-appendix1}, a similar reasoning is applied to the case of the gradient with respect to the initial condition.
\begin{proposition}[gradient dynamics -- $x_0$]
\label{prop2} Let
\begin{equation}
    \Lambda_{0,k} \doteq \frac{\d x_k}{\d {x}_{0}} \in \mathbb R^{n_x,n_x}.
    \label{eqn:phi0def}
\end{equation}
be the matrix containing the total derivatives of the states with respect to $x_0$. Consider \eqref{eqn:rho-var-def}.
The gradient evolution with respect to the initial condition along the multi-step horizon is obtained by means of the following time-varying dynamical system
\begin{subequations}
    \begin{align}
        \label{eqn:GUlaw_initcond_mat}
         &\Lambda_{0,k} = \mathcal{J}^{x\!/\!x}_k\Lambda_{0,k-1},\\
       \label{eqn:GUlaw_initcond}
         &\nabla_{x_0} \mathcal C_{k} = \nabla_{x_0} \mathcal C_{k-1} + {\Lambda^\top_{0,k}}\rho_k,
    \end{align} 
    \label{GULAW_x0}
    \end{subequations}
with $\Lambda_{0,0} \doteq \frac{\d x_0}{\d x_0} = \mathbb{I}_{n_x}$, 
$\nabla_{x_0} \mathcal C_{0} = {\Lambda^\top_{0,0}}\rho_0$.
\end{proposition}
In this case, it is noticeable that there is no ``direct" effect of $\widehat x_0$ on~$\widehat{x}_k$, as the impact of the estimated initial condition affects the predictions only through its propagation over time. This is shown by the term $\Lambda_{0,k-1}$. Similarly to $\Lambda_{k-1}$, it acts as a ``memory" that captures how the effect of $\widehat x_0$ on past predictions, i.e.,~$\widehat x_\tau$ with $\tau\in[1,k-1]$, has influenced the current state estimation $\widehat x_k$.

The extension of Proposition~\ref{prop1} and Proposition~\ref{prop2} to the cost function \eqref{eqn:final_cost} including physical penalties and black-box regularization is reported in the following remark.
\begin{remark}[extension to the overall framework]\label{rmk:overall_framework}
    When physics-based penalty terms and the black-box regularization are simultaneously introduced in the cost function $\mathcal{C}_T$, resulting in \eqref{eqn:final_cost},
%
it is possible to recursively update the gradients using \eqref{eqn:GULAW}, and \eqref{GULAW_x0} by introducing a modification in the two terms defined in \eqref{eqn:rho-var-def} as follows
$$
\begin{gathered}
    \begin{aligned}
        \rho_k\!\doteq\! \left({\nabla^\top_{e} \mathcal{L}_k}{\mathcal{J}^{e\!/\!z}_k}{\mathcal{J}^{z\!/\!x}_k}\right)^\top\!+\!
        \lambda{\nabla_{x} p}(\widehat x_k, \theta)\!+\!\nu{\nabla_{x} q}^2(\widehat x_k, \theta) 
        ,
    \end{aligned}
    \label{eqn:rhodef_phy}
    \\
    \begin{aligned}
        \varrho_k\!\doteq\!{\nabla_{\vartheta} \mathcal{L}_k}\! + \!\lambda{\nabla_{\vartheta} p}(\widehat x_k,\!\theta)\!+\!\nu{\nabla_{\vartheta} q}^2(\widehat x_k,\!\theta)\!+\!\rho{\nabla_{\vartheta} \|\Omega\|_{1\beta}},
    \end{aligned}
    \label{eqn:vardef_phy}
\end{gathered}
$$
considering the extended vector of parameters $\vartheta = [\theta^\top,\mathrm{vec}(\omega)^\top]^\top \in \mathbb R^{n_\vartheta}$, i.e., the column vector containing all the physical and black-box parameters.
\end{remark}
\subsection{Systems identifiability}\label{sec:identifyability}
In general, the uniqueness of the solutions of first-order methods depends on the identifiability of the system being analyzed. Specifically, the obtained solution may correspond to a unique local minimum of the cost function or, alternatively, a plateau region. The concept of identifiability, which plays a crucial role in determining convergence behavior, is formally discussed next following the definitions in \cite{bellman1970structural}.
\begin{remark}[system identifiability]
    It must be remarked that not all the realizations of the system $\mathcal{S}$ are identifiable, meaning that (locally) the solution may not be unique. Indeed, a specific system $\mathcal{S}$ with parameters $\bar \theta$ and initial conditions $\bar{x}_0$ is said to be (locally) identifiable if the cost function $\mathcal{C}_T$ has a strict local minimum at $\widehat \theta = \bar \theta$, $\hxo = \bar x_0$.
    If the minimum is global, the structure is said to be globally identifiable.
    A sufficient condition for the structure $\mathcal{S}$ to be locally identifiable is that the Hessian matrix with respect to $\widehat{\theta}$ and $\hxo$, is positive definite for all $\theta \in \Theta$, $x \in \mathcal{X}$, being $\Theta$ and $\mathcal{X}$ suitable neighborhoods of $\bar \theta$, $\bar x_0$.
    \label{rmk:identifiability}
\end{remark}

\subsection{Computational efficiency}

The proposed approach offers a more efficient way to compute the gradients compared to the analytic formula proposed in \cite{donati2023oneshot}, as stated in the following remark. 
\begin{remark}[complexity improvement]
    Let $T$ represent the length of the prediction horizon considered in the system identification process. It can be shown that the computational complexity of the gradient computation algorithm proposed in \cite{donati2023oneshot} scales as $\mathcal{O}(T^3n_x^2n_\theta)$, making it impractical for large-scale problems due to its cubic growth rate. In contrast, the dynamical system for the gradient that we propose in this section scales linearly with the length of the horizon, exhibiting a computational complexity of $\mathcal{O} (Tn_x^2n_\theta)$.
\end{remark}

To {illustrate thus improvement}, we briefly revisit the same numerical example proposed in \cite{donati2023oneshot}, In which the attitude dynamics of the satellite {was} modeled using the Euler equations. As shown in Table \ref{tab:comp_rec}, by dynamically computing the impact of $\widehat \theta$ and $\widehat x_0$ time-step by time-step and utilizing the memory matrix $\Lambda$ along with predictions, the computational efficiency is significantly improved, making the gradient calculation both faster, and consequently significantly more scalable, thus leading to lower training times and improved performance.

\begin{table}[!ht]
    \centering
    \caption{Average gradient calculation time (seconds) for $T~=~50, 100, 150, 200, 250$.}
    \begin{tabular}{c c c c c c c}
    \hline
         T & 50 & 100 & 150 & 200 & 250\\
         \hline
         Rec. BP. & $0.0015$ & $0.0020$ & $0.0035$ & $0.0043$ & $0.0049$\\
         An. BP.  & $0.0411$ & $0.2877$ & $0.9300$&  $2.3048$& $4.2128$\\
     \hline
    \end{tabular}
    \label{tab:comp_rec}
\end{table}

\subsection{The proposed approach}
In this section, we present in Algorithm \ref{alg:algGD} the proposed identification approach, considering physical penalties and black-box compensation (see Remark \ref{rmk:overall_framework}). We propagate the model \eqref{eqn:bb_extension_v2} with initial conditions $\widehat x_0$ and parameters $\widehat\theta$ and $\omega$, while updating the gradient according to Proposition \ref{prop1} and \ref{prop2} along the horizon $T$. Then, accordingly, we update the weights. 
This process repeats until at least one of the following conditions is satisfied: (a) the structure converges to a (possibly local) minimum of the loss function, or below a given threshold $\varepsilon_1$; (b) the magnitude of the gradient is lower than a given minimum step size $\varepsilon_2$.
\begin{algorithm}
\caption{The Identification algorithm}\label{alg:algGD}
\begin{algorithmic}[1]
\State Given input-output observations over an horizon of length $T$, $\{\mathbf{\widetilde u}_{0:T},\mathbf{\widetilde z}_{0:T}\}$, and a dictionary of basis functions $\varphi$, choose $\alpha_\vartheta$, $\alpha_{x_0}$, and $\varepsilon_1$, $\varepsilon_2$ sufficiently small.
\State {Initialize} $\widehat{x}_0$, $\widehat\vartheta = [\widehat\theta^{\top},\mathrm{vec}(\omega)^{\top}]^\top$.
\While{$\mathcal{C}_T \geq \varepsilon_1$ \textbf{and} $\|\nabla\mathcal{C}_T\|_2 \geq \varepsilon_2$ }
\State {Initialize} $k = 0$, $\Lambda_{0} = \mathbf{0}_{n_x,n_\theta}$, $\nabla_{{\vartheta}} \mathcal C_{0} = \mathbf{0}_{n_\vartheta}$, $\Lambda_{0,0} = \mathbb{I}_{n_x}$, and $\nabla_{x_0} \mathcal C_{0} = {\Lambda^\top_{0,0}}\rho_0$.
\While{ $k \leq T$ }
\State Predict $\widehat x_{k+1}$, $\widehat z_k$ using \eqref{eqn:bb_extension_v2} with $\widehat{x}_0$, $\widehat{\vartheta}$.
\State Compute ${e}_{k} = \widehat z_k-\widetilde z_k$ and $\mathcal{L}_k$.
\State Compute $\Lambda_{k}$ \eqref{eqn:GUlaw_mat} and $\Lambda_{0,k}$ \eqref{eqn:GUlaw_initcond_mat}.
\State Update the gradients, i.e., compute  $\nabla_{{\vartheta}} \mathcal C_{k}$ and $\nabla_{x_0} \mathcal C_{k}$ using  \eqref{eqn:GUlaw_grad} and \eqref{eqn:GUlaw_initcond}.
\State $k \gets k+1$
\EndWhile
\State Compute the final cost $\mathcal{C}_T$ \eqref{eqn:final_cost}.
\State $\nabla\mathcal{C}_T = [\nabla^\top_\vartheta\mathcal{C}_T, \nabla^\top_{x_0}\mathcal{C}_T]^\top$.
\State Update the weights using any first-order method, e.g., gradient descent \eqref{eqn:GD}.
\EndWhile
\State Return $\vartheta^\star=\widehat\vartheta$ and ${x}_0^\star=\widehat{x}_0$
\end{algorithmic}
\end{algorithm}

\section{Theoretical properties}\label{sec:theo}
We propose a thorough theoretical analysis of the proposed approach, presenting key theoretical properties, including sufficient conditions for the stability of the gradient and bounds on the parameter estimation error. Clearly, due to the nonlinearity and nonconvexity of the problem, most of the presented results have a \textit{local} nature, {i.e., they hold when Algorithm 1 is initialized sufficiently close to the optimal solution}.

\subsection{Non-exploding gradient}\label{sec:explgrad}
In this section, we exploit the dynamical formulation \eqref{eqn:GULAW} to obtain sufficient conditions for the stability of the gradient with respect to the parameter $\theta$, useful to ensure that gradients do not grow unboundedly\footnote{We remark that the same reasoning with analogous results applies also to the gradient for initial condition, which is not reported here for space reasons.}. 
The uniform Bounded-Input, Bounded-Output (BIBO) stability for linear time systems is exploited in the following theorem. This concept, formally expressed in \cite{rugh1996linear}, is reported in the following definition.
\begin{definition}[uniform BIBO stability]
    A linear state equation is called uniformly BIBO stable if there exists a finite constant $\eta$ such that for any $k_0$ and any input signal $u_k$ the corresponding zero-state response $y_k$ satisfies
    $$
    \sup_{k\geq k_0} \|y_k\| \leq \eta \sup_{k\geq k_0} \|u_k\|
    $$
\end{definition}
Then, the concept of non-exploding (or unbounded) gradient is formally defined as follows.
\begin{definition}[non-exploding gradient]
    The multi-step gradient $\nabla_{{\theta}} \mathcal C_{k}$ is said to be non-exploding (or bounded) if and only if there exists a finite constant $\eta$ such that for any $k_0$ satisfies
    $$
    \sup_{k\geq k_0} \|\nabla_{{\theta}} \mathcal C_{k}\| \leq \eta
    $$
\end{definition}
Thus, the following theorem formalizes the link between the boundedness of multi-step gradients and the BIBO stability of the corresponding LTV system \eqref{eqn:GULAW}.
\begin{theorem}[gradient stability] \label{thm:exp.grad}
    The multi-step gradient $\nabla_{{\theta}} \mathcal C_{k}$ is non-exploding if and only if the associated LTV system is BIBO stable for the specific sequence of inputs $u_k = 1, \forall k$, i.e., there exists a finite constant $p$ such that
    \begin{equation}
    \sum_{i=j}^{k-1}\|\Lambda^\top_i\rho_i+ \varrho_i\|\leq p
    \label{eqn:theo2cond}
    \end{equation}
    for all $k \in [1, T],j$ with $k\geq j+1$.
\end{theorem}
Proof of Theorem \ref{thm:exp.grad} is reported in Appendix \ref{app:exp.grad}.

Consequently, certain conditions on the system's predicted trajectory can provide sufficient guarantees as stated in the following corollary, proved and discussed in Appendix \ref{app:corollary.exp.grad}.
\begin{corollary}[trajectory to gradient stability]\label{cor:corollary.exp.grad}
    For a specific current value of $\widehat \theta \in \mathbb R^{n_\theta}$, a sufficient condition for the gradient to be non-exploding is that the predicted trajectory defined by the sequence of states $\{x_k\}_{k=0}^{T}$ satisfying the system’s dynamics has not a finite escape time $k_c < T$, such that
    \begin{equation*}
    \lim_{k \to k_c} \lVert x_k \rVert_p = \infty,
    \end{equation*}
   for suitable $\ell_p$ norm, and $k_c$ is the critical time step at which the trajectory diverges to infinity, i.e., it is not \textit{finite-time unstable} with $k_c < T$.
\end{corollary}
\begin{remark}[unstable trajectories]
    The identification with unstable, but not in finite time, trajectories, can still be handled with a proper adaptation of the learning rate.
\end{remark}

\begin{remark}[exploding gradient alternatives]
    The problem of exploding gradient can also be handled by incorporating  techniques such as gradient clipping or truncated gradient in the defined gradient updated system.
\end{remark}

\subsection{Estimation error bound}\label{sec:bound}
In this subsection, we compute an upper bound on the parametric identification error, i.e., the maximum distance between the {{global}} minimizer of the cost function $\theta^\star$ and the true parameters vector $\bar \theta$. {{Clearly, the optimization problem will converge to the global minimizer only if it is initialized sufficiently close to its region of attraction}}. For simplicity of exposition, we consider the case when the initial condition $x_0$ is known, and we only refer to the vector of physical parameters $\theta$. The extension to the general case is straightforward. 
%

First, we observe that, as discussed in Section \ref{sec-OW}, the cost $\mathcal{C}_{T}$ can be seen as a function of the noise sequences and the parameter vector (see equation \eqref{eq:CT_eta}), that is, we write $\mathcal{C}_{T}(\theta;\boldsymbol{\eta})$, with $\boldsymbol{\eta} \doteq \{\boldsymbol{\eta}^{u}_{0:T}, \boldsymbol{\eta}^{z}_{0:T}\}$.
Then, to derive the parametric error bound, we consider the following assumptions.
\begin{assumption}[preliminary assumptions]\label{assumpt:general}~
    \begin{enumerate}[i)]
\item \label{assumpt:general.id}(local identifiability)
The Hessian of the loss function evaluated in $\theta^\star$ is positive definite for all possible values of the noise, i.e.,
\[
H\doteq\left.\frac{\partial^{2}\mathcal{C}_{T}(\theta;\boldsymbol{\eta})}{\partial^{2}\theta}\right|_{\theta=\theta^{\star}}\succ0, 
\quad \forall \boldsymbol{\eta}
\]
\item \label{assumpt:general.local}(convergence to $\bar \theta$) When the noise is null and~${\Delta = 0}$, minimization of the loss function gives the
true parameter vector: $\theta^{\star}=\bar{\theta}$.
\end{enumerate}
\end{assumption}
Assumption \ref{assumpt:general}.\ref{assumpt:general.id}) is reasonable and standard in this context \cite{milanese1991optimal,Ozay2012bounded}
Assumption \textit{\ref{assumpt:general}.\ref{assumpt:general.local})} is certainly satisfied if the initial point of the minimization algorithm is in the region of attraction of the true parameter vector $\bar{\theta}$. Indeed, if $\Delta=0$, the noise is null and the identifiability assumption holds, we have that $\mathcal{C}_T(\bar{\theta}; \boldsymbol{0}) = 0$, implying that
the problem local solution is $\theta^\star=\bar{\theta}$.

Thus, we state the following lemma.
\begin{lemma}[bounded function] \label{lemma:bounded}
    Let Assumption \ref{assumpt:general} hold. Assume $\boldsymbol{\eta} \in \mathcal{N}$ with $\mathcal{N}$ {closed and bounded}. Define
\[
G(\boldsymbol{\eta})\doteq\left.\frac{\partial^{2}\mathcal{C}_{T}(\theta;\boldsymbol{\eta})}{\partial\boldsymbol{\eta}\partial\theta}\right|_{\theta=\theta^{\star}}.
\]
The function $H^{-1}G(\boldsymbol{\eta})$ is bounded for all $\boldsymbol{\eta}\in \mathcal{N}$.
In particular, two constants $M_{u},M_{z}<\infty$ exist, such that
\[
\begin{alignedat}{1} & \max_{\boldsymbol{\eta}\in \mathcal{N}}\left\Vert \left[H^{-1}G(\boldsymbol{\eta})\right]_{u}\right\Vert _{p}<M_{u}\\
 & \max_{\boldsymbol{\eta}\in \mathcal{N}}\left\Vert \left[H^{-1}G(\boldsymbol{\eta})\right]_{z}\right\Vert _{p}<M_{z}
\end{alignedat}
\]
where $\left\Vert \cdot\right\Vert _{p}$ is the matrix $p$-norm,
and $\left[H^{-1}G\right]_{u}$ and $\left[H^{-1}G\right]_{z}$ are
the matrices containing the columns of $H^{-1}G$ corresponding to
$\boldsymbol{\eta}^{u}_{0:T}$ and $\boldsymbol{\eta}^{z}_{0:T}$, respectively.
\end{lemma}
\noindent\textit{Proof}$\quad$ We have from Assumption \textit{\ref{assumpt:general}.\ref{assumpt:general.id})} that $H$ is positive definite $\forall \boldsymbol{\eta} \in \mathcal{N}$. Thus, it follows that it is invertible and the inverse is bounded. Moreover, we have that $\mathcal{C}_T$ is twice continuously differentiable since it is defined as a sum of twice differentiable functions with continuous derivatives $\mathcal{L}_k,$ $\forall k \in [0, T]$. It follows from Lipschitz continuity that $G(\boldsymbol{\eta})\doteq\left.\frac{\partial^{2}\mathcal{C}_{T}(\theta;\boldsymbol{\eta})}{\partial\boldsymbol{\eta}\partial\theta}\right|_{\theta=\theta^{\star}}$ is bounded. The statement of the lemma follows considering that $\mathcal{N}$ is {closed and bounded}. \hfill $\blacksquare$

Notice that, the values of $M_u$ and $M_z$ can be estimated using, e.g., Monte Carlo or probabilistic methods \cite{tempo2013randomized}.
%
Thus, the following theorem on the bound of the parametric error can be stated.
\begin{theorem}[{parametric error bound}] \label{thm:bounds} Let Assumption \ref{assumpt:general} holds. Then, the {parametric identification error} is bounded as
\[
\left\Vert \theta^{\star}-\bar{\theta}\right\Vert _{p}\leq M_{u}\left\Vert \boldsymbol{\eta}^{u}_{0:T}\right\Vert_p+M_{z}\left\Vert \boldsymbol{\eta}^{z}_{0:T}\right\Vert_p + M_{\Delta}\|\boldsymbol{\boldsymbol{\widetilde \Delta}}\|_p.
\]
where $\boldsymbol{\boldsymbol{\widetilde \Delta}} = \{\widetilde \Delta_1,\dots, \widetilde \Delta_T\}$ with $\widetilde \Delta_k \doteq \Delta(x_k, u_k) - \delta(\widehat x_k, u_k; \omega)$ is the sequence of residual terms not compensated by the black model $\delta(\cdot)$, and a constant $M_{\Delta}<\infty$. 
\end{theorem}
Proof of Theorem \ref{thm:bounds} is reported in Appendix \ref{app:bounds}.
\begin{remark}[well-defined identification problems]
    Since the parametric identification error is inversely proportional to $H^{-1}$, a direct consequence of Lemma~\ref{lemma:bounded} and Theorem \ref{thm:bounds} is that a well-defined identification problem, characterized by a ``large'' invertible Hessian matrix $H$ (see Remark \ref{rmk:identifiability}), guarantees a ``small'' parametric identification error, thereby ensuring an accurate representation of the true system dynamics through the estimated parameters.
\end{remark}

\begin{remark}[black-box compensation effect]
From Theorem \ref{thm:bounds} it can be also observed that as the black model $\delta(\cdot)$ more effectively compensates the effect of $\Delta(\cdot)$, that is, as $\widetilde\Delta_k \rightarrow 0,\, \forall k$, the parametric error becomes more tightly bound. This implies that an efficient compensation by $\delta(\cdot)$ leads to a reduction in the upper bounds of the parametric error, enhancing the accuracy of the physical parameters estimation (see Subsection \ref{subsec:opt_err}).
\end{remark}
{In a typical situation, we do not know the unmodeled term $\Delta(x_k,u_k)$, but we know some information about it, like an upper bound on its norm. Moreover, it is reasonable to assume the noises to be bounded with known bounds, i.e., \begin{equation}
\boldsymbol{\eta}\in \mathcal{N}\doteq\{{\boldsymbol{\eta}}:\left\Vert \boldsymbol{\eta}^{u}_{0:T}\right\Vert _{p}\leq\bar{\eta}_{u},\left\Vert \boldsymbol{\eta}^{z}_{0:T}\right\Vert _{p}\leq\bar{\eta}_{z}\}.\label{eq:noiseb}
\end{equation}
Thus, assuming that an upper bound $\|\boldsymbol{\widetilde \Delta}\|_p \leq \bar \Delta$ on the residuals sequence can be estimated, it is immediate to prove the following corollary of Theorem~\ref{thm:bounds}.
\begin{corollary}[bounded residuals]
    Let Assumption \ref{assumpt:general} holds. Let $\boldsymbol{\eta}\in \mathcal{N}$ and $\|\boldsymbol{\widetilde \Delta}\|_p \leq \bar \Delta$. Then, the {parametric identification error} is bounded as
    \[
    \left\Vert \theta^{\star}-\bar{\theta}\right\Vert _{p}\leq M_{u}\bar{\eta}_{u}+M_{z}\bar{\eta}_{z} + M_{\Delta}\bar \Delta.
    \]
\end{corollary}

\section{Maximum sparsity recovery}
\label{sec:sparsityrecovery}
In this section, we provide a theoretical analysis on the recovery of the maximum sparsity of the black-box contribution. First, we analyze a simpler problem and present results for the case of linear-in-parameters and single-step models. Subsequently, we extend the obtained results to the more general case of multi-step, and possibly nonlinear-in-parameters models.

\begin{remark}[black-box regularization]
    The objective of trying to find the sparsest representation of the unmodelled dynamics is to have a description that it is easier to analyze and interpret. However, this is not the only criterion that one can use to choose the ``right'' black-box model. In many applications, one may want to describe the data mainly by the grey-box part of the model and, hence, it is desirable to find the ``smallest'' black-box that (together with the grey-box) explains the data available. This can be done by minimizing any norm of the black-box part. 
\end{remark}

\subsection{Single-step, linear-in-parameters setting}
Consider a dynamical system described by {the state-equation representation} \eqref{eqn:system}, with parameters $\bar \theta \in \mathbb R^{n_\theta}$, having a linear-in-parameters function $f$, full state observations, and, without loss of generality, $n_x=n_z=1$. 
In such a scenario, $\mathcal S$ takes the form
\begin{equation}
    \mathcal{S}:\quad x_{k+1} =  {\xi}^\top(x_k,u_k) \bar \theta + \Delta(x_k,u_k),
\label{eqn:system.bar}
\end{equation}
with {$z_k = x_k$, and}
$\xi(x_k,u_k):\mathbb R^{n_x} \times \mathbb R^{n_u} \rightarrow \mathbb R^{n_\theta}$ a vector of known (possibly nonlinear) physical terms defined as
$\xi(x_k,u_k) = [\xi_1(x_k,u_k),\dots,\xi_{n_\theta}(x_k,u_k)]^\top$.
A dictionary of basis functions $\varphi \in \mathbb R^{m}$ and
a set of noise-corrupted data~${\mathcal{D}=\{\mathbf{{\widetilde u}}_{0:T}, \mathbf{{\widetilde z}}_{0:T}\}}$, collected from \eqref{eqn:system.bar}, are available. In this case, the relationship between $\widetilde u_k$ and $\widetilde z_k$ is described by
$$
\widetilde z_{k+1} = {\xi}^\top(x_k,\widetilde u_k) \bar \theta + \Delta(x_k,\widetilde u_k) + \eta^v_k
$$
where $\eta^v_k$ is a noise term accounting for both the process and measurement noises $\eta^u_k$ and $\eta^z_k$, as detailed in the measurement model described in Appendix \ref{app:measurement_model}.

\begin{assumption}[unknown but bounded noise]
The noise sequence $\eta^v=[\eta^v_1,\dots,\eta^v_T]^\top$ is assumed to be unknown but bounded, i.e., $\|\eta^v\|_2 \leq \mu$.
\label{ass:noise}
\end{assumption}
Hence, let us define the matrices~${\Xi\in \mathbb R^{T,n_\theta}}$ and~${\Phi \in \mathbb R^{T,m}}$ as follows 
$$
\begin{aligned}
    \Xi &\doteq \left[\begin{array}{ccc}
     \xi_1(\widetilde x_0, \widetilde u_0)&\dots& \xi_{n_\theta}(\widetilde x_0, \widetilde u_0)\\
     \vdots&\ddots&\vdots\\
     \xi_1(\widetilde x_{T-1}, \widetilde u_{T-1})&\dots& \xi_{n_\theta}(\widetilde x_{T-1}, \widetilde u_{T-1})
\end{array}\right]\\
\end{aligned}
$$
$$\begin{aligned}
    \Phi &\doteq \left[\begin{array}{ccc}
             \varphi_1(\widetilde x_0, \widetilde u_0)&\dots& \varphi_m(\widetilde x_0, \widetilde u_0)\\
             \vdots&\ddots&\vdots\\
             \varphi_1(\widetilde x_{T-1},\widetilde u_{T-1})&\dots& \varphi_m(\widetilde x_{T-1}, \widetilde u_{T-1})
        \end{array}\right]\\
        \end{aligned}
        $$
where $\widetilde x_k \doteq \widetilde z_k$, since we are initially considering the states to be fully measured \eqref{eqn:system.bar}.
%
%
\subsection{Problem definition}
We address the problem of finding a sparse linear combination of the basis functions in the dictionary that, combined with the prior physical model, is consistent with the measured data. 
First, we formally state the definitions of feasible and maximally sparse coefficients.
\begin{definition}[feasible coefficients]
\label{def:feasiblecoeff}
    Given two vectors, $\omega_0 \in \mathbb R^m$ and~${\theta_0 \in \mathbb R^{n_\theta}}$, a set of noise-corrupted data $\mathcal{D}=\{\mathbf{{\widetilde u}}_{0:T}, \mathbf{{\widetilde z}}_{0:T}\}$ satisfying Assumption \ref{ass:noise}, a sequence of states (predicted or measured)~$\mathbf{x}_{0:T-1}$, and the bound on the noise $\mu$, we say that $\omega_0$ is \textit{feasible} for $\theta_0$ if and only if~${\omega_0 \in FPS(\theta_0)}$, where  
    $$\begin{aligned}
    &FPS(\theta_0) \\&\,\,\doteq \left\{ \omega \in \mathbb R^m: \|\widetilde z\!-\!f(\mathbf{ x}_{0:T-1}, \mathbf{\widetilde u}_{0,T-1}, \theta_0)\!-\!\Phi \omega\|_2 \leq \mu \right\}.
    \end{aligned}$$
    with $\widetilde z = [\widetilde z_1,\dots, \widetilde z_T]^\top.$
    More in general, a vector $\omega \in \mathbb R^m$ is feasible if and only if 
     $\exists \theta \in \mathbb R^{n_\theta} : \omega \in FPS(\theta).$
\end{definition}
\begin{definition}[maximally sparse coefficients]
Given a set of noise-corrupted data $\mathcal{D}=\{\mathbf{{\widetilde u}}_{0:T}, \mathbf{{\widetilde z}}_{0:T}\}$ satisfying Assumption \ref{ass:noise}, a sequence of states (predicted or measured)~$\mathbf{x}_{0:T-1}$, and the bound on the noise~$\mu$,
a feasible coefficient vector is said \emph{maximally sparse} if it is a solution of the optimization problem 
\begin{equation}
    \begin{array}{cl}
        \bar \omega\!=\!&\arg\min\limits_{\theta,\omega} \|\omega\|_0\\
        & \mathrm{{s.t.}}\; \|\widetilde z\!-\!f(\mathbf{x}_{0:T-1}, \mathbf{\widetilde u}_{0,T-1}, \theta)\!-\!\Phi \omega\|_2\!\leq\! \mu
    \end{array}
    \label{eqn:opt_prob}
\end{equation}
where $\|\cdot\|_0$ is the $\ell_0$ quasi-norm.
\end{definition}
Hence, we formally introduce the simplified sparsity problem, discussed in the first part of this analysis.
\begin{problem}[simplified sparsity problem] \label{prb:A}
Defining the following parametrized, single-step, state-equation model
\begin{equation*}
\begin{aligned}
    \mathcal{M}:\quad \widehat x_{k+1} &= \sum_{i=1}^{n_\theta }\theta_i\xi_i(\widetilde x_k,\widetilde u_k) + \sum_{i=1}^m \omega_i\varphi_i(\widetilde x_k, \widetilde u_k),\\
    &= \widehat F(\widetilde x_k,\widetilde u_k,\theta,\omega) \\
    \widehat z_k &= \widehat x_k,
    \end{aligned}
\end{equation*}
the goal is to identify approximations $\widehat \theta, \widehat \omega$ of $\theta, \omega$ from the data set $\mathcal D$, such that:
\begin{enumerate}[i)]
    \item  $\widehat \omega$ is sparse;
    \item $(\widehat \theta, \widehat \omega)$ are consistent with the dataset, i.e., the single-step prediction error satisfies $$\|\widetilde z~-~\widehat F(\mathbf{\widetilde x}_{0:T-1},\mathbf{\widetilde u}_{0:T-1}, \widehat{\theta}, \widehat{\omega})\|_2 \leq \mu.$$
\end{enumerate}
\end{problem}
\begin{assumption}[problem feasibility]
    Considering the dataset $\mathcal{D}$, we assume that there exists at least 
    one feasible $\omega$, according to Definition \ref{def:feasiblecoeff}. 
    \label{ass:feasibleprob}
\end{assumption}
{Note that no assumptions are made on the structural form of the black-box term $\Delta$ in \eqref{eqn:system.bar}. We only suppose that the chosen dictionary of basis function $\varphi$ is sufficiently rich to allow an approximation of $\Delta$ compatible with the given noise level.}

Obviously, a solution to the sparse identification Problem \ref{prb:A} could be obtained by solving the single-step optimization problem \eqref{eqn:opt_prob}, where the sequence of states measurements $\widetilde{\mathbf{x}}_{0:T-1}$ is exploited to perform the single-step predictions.
However, it cannot be solved in general, since the quasi-norm is a nonconvex function and its minimization is an NP-hard problem. Instead, we aim to solve the following relaxed version, i.e.,
\begin{equation}
    \begin{array}{cl}
        (\widehat \theta, \widehat \omega) = &\arg \min\limits_{\theta, \omega} \|\omega\|_1\\
        & \text{s.t. } \|{\widetilde z}\!-\!\widehat F(\mathbf{\widetilde x}_{0:T-1},\mathbf{\widetilde u}_{0:T-1},\theta,\omega)\|_2\!\leq\!\mu.
    \end{array}
    \label{eqn:optprob1}
\end{equation}
The optimization problem \eqref{eqn:optprob1} is a convex $\ell_1$-relaxed version of problem \eqref{eqn:opt_prob}, proposed for solving the sparse identification Problem \eqref{prb:A}. 
\subsection{Theoretical analysis}
In this section, we present two theorems based on the results in \cite{gribonval2006simple}, \cite{novara2012sparse}. These theorems give conditions under which a coefficient vector is maximally sparse. Hence, such a vector ensures the selection of the minimum number of basis functions in the dictionary $\varphi$ to compensate for model uncertainties, while preserving consistency with the dataset. Here, we extend these two theorems to scenarios where only a subset of the parameters requires sparsification. Specifically, in our case, we aim to regularize only the coefficients associated with the black-box component. Before stating the theorems we introduce a few notations in the following definition.
\begin{definition}[preliminary notations]
    For each integer $n \in \mathbb R$ and matrix $Q \in \mathbb R^{n_1,n_2}$ we denote 
$$\sigma^2_{min,n}(Q) \doteq \inf_{\|x\|_0\leq n} \frac{\|Qx\|^2_2}{\|x\|_2^2} \leq 1.$$
Moreover, the following norm is defined
$$\|x\|_{(Q,n)} \doteq \sqrt{\sum_{i\in \mathcal{I}_n(x)} \left(x^\top q_i\right)^2},$$
where $\mathcal{I}_n(x)$ indexes the $n$ largest inner products $|x^\top q_i|$ with $q_i$ the $i$-th column of $Q$.
\end{definition}
%
Then, we define the prediction error, representing the discrepancy between measurements and predictions obtained when the black-box augmentation is employed. Thus, in the subsequent lemma, we compute the best-case compensation error.
\begin{definition}[prediction error]
Given a generic observations vector of length $T$, i.e., $z \in \mathbb R^{T}$, and two matrices $P\in \mathbb R^{T,n_\theta}$, $Q\in \mathbb R^{T,m}$ such that~$z = P\theta + Q\omega$,
define the prediction error as follows
\begin{equation}
    e_{P,Q}(z, \theta, \omega) = z - P\theta - Q\omega,
    \label{eqn:pred_error}
\end{equation}
such that, e.g., $e_{\Xi,\Phi}(\widetilde z, \theta, \omega) = \widetilde z- \Xi\theta - \Phi\omega$.
\end{definition}
\begin{lemma}
[optimal compensation error]
    Given $P\in \mathbb R^{T,n_\theta}$, $Q \in \mathbb R^{T,m}$, and $z \in \mathbb R^{T}$, such that~$z = P\theta + Q\omega$ define 
    \begin{equation*}
        \Upsilon(P) = (I_T - P P^\dag),
    \end{equation*}
    where $\dag$ denotes the pseudo-inverse operator.
    Considering \eqref{eqn:pred_error}, {the optimal compensation error} is
    \begin{equation*}
    \begin{aligned}
    {e}_{P,Q}^{\star}(z,\omega) &\doteq e_{P,Q}\left(z,\arg \min_{\theta \in \mathbb R^{n_\theta}} \|e_{P,Q}(\theta,\omega)\|_2, \omega\right)\\ &= \Upsilon(P)( z - Q \omega).
    \end{aligned}
    \end{equation*}
\end{lemma}
\noindent\textit{Proof}$\quad$ Consider the problem $$\min_{\theta \in \mathbb R^{n_\theta}} \|e_{P,Q}(z,\omega, \theta)\|_2 = \|(z - Q\omega ) - P\theta\|_2,$$
for which the optimum is achieved at the least squares solution, i.e.,
$$\theta^\star(z,\omega) = (P^\top P)^{-1}P^\top({ z} -Q\omega) = P^\dag({ z} -Q\omega).$$ Thus,
$$
\begin{aligned}
e_{P,Q}(z,\theta^\star(\omega), \omega) &=z\!-\!PP^\dag({ z} -Q\omega) - Q\omega\\ &= (I_T - PP^\dag) z- (I_T - PP^\dag) Q\omega. \hfill \blacksquare
\end{aligned}
$$

Hence, the following theorems can be stated, providing conditions to check if an estimation $\widehat\omega$, e.g., the solution of \eqref{eqn:optprob1}, has the same support as $\bar \omega$, thus is maximally sparse as well.
First, we use the results of \cite{gribonval2006simple} to formally define the conditions under which two vectors have equivalent supports.
\begin{theorem}[equivalent support conditions] \label{thm:grib_var}
Let~$\widehat \omega \in \mathbb R^m$ be the solution of the optimization problem \eqref{eqn:optprob1}, ${M} \doteq \| \widehat\omega\|_0$ the cardinality of its support, and~${\Upsilon_0 = \Upsilon(\Xi)}$. Let $\omega'$ be any other representation, and~${\widetilde z = [\widetilde z_1,\dots, \widetilde z_T]^\top}$ the vector of noise corrupted observations collected from \eqref{eqn:system.bar}.
If the following inequalities hold
    \begin{subequations}
    \begin{align*}
        \|e^{\star}_{\Xi,\Phi}(\widetilde z, \omega')\|_2 &\leq \|e^{\star}_{\Xi,\Phi}(\widetilde z, \widehat \omega)\|_2,\\
        \|\omega'\|_0 &\leq M,
    \end{align*}
    \end{subequations}
    then
    \begin{equation*}
        \|\omega'-\widehat\omega\|_\infty \leq \frac{\|e^{\star}_{\Xi,\Phi}(\widetilde z, \widehat\omega)\|_{(\Upsilon_0,1)} + \|e^{\star}_{\Xi,\Phi}(\widetilde z, \widehat\omega)\|_{(\Upsilon_0,2M)}}{\sigma_{min,2M}^2(\Upsilon_0)}.
    \end{equation*}
    Moreover, if
    \begin{equation*}
    \begin{aligned}
        \|e^{\star}_{\Xi,\Phi}(\widetilde z, \widehat\omega)\|_{(\Upsilon_0,1)} &+ \|e^{\star}_{\Xi,\Phi}(\widetilde z, \widehat\omega)\|_{(\Upsilon_0,2M)}\\ &< \sigma_{min,2M}^2 (\Upsilon_0)\eta(\widehat \omega),
    \end{aligned}
    \end{equation*}
    with $\eta(\omega) \doteq \min\limits_{i\in\text{supp}(\omega)} |\omega_i|$,
    then $\widehat\omega$ and $\omega'$ have the same support,
    $$\mathrm{supp}(\widehat\omega) = \mathrm{supp}(\omega'),$$ and the same sign, $$\mathrm{sign}(\widehat\omega_i) = \mathrm{sign}(\omega'_i), \quad \forall i\in[1,m].$$
\end{theorem}  
The proof of Theorem \ref{thm:grib_var} is reported in Appendix \ref{app:grib_var}
\begin{remark}[generality of the sparsity conditions]
    While Theorem \ref{thm:grib_var} is specific and shows how to apply the results of \cite{gribonval2006simple} to the solution of \eqref{eqn:optprob1}, it must be remarked that its conditions can be verified on any other vector $\omega \in \mathbb R^m$ such that~$\Upsilon_0 \widetilde z=\Upsilon_0\Phi\omega+e^{\star}_{\Xi,\Phi}(\widetilde z,\omega)$.
\end{remark}
Then, based on the results proposed in \cite{novara2012sparse}, the following theorem formally defines the conditions for which a vector is maximally sparse.
Let us first define the vector $\omega^v$, solution of the following optimization problem 
\begin{subequations}
    \begin{align}
        \omega^v=&\arg \min\limits_{ \omega} \|\omega\|_1\\
        & \text{s.t. } \,\,\,\omega_i \geq \text{sign}(\widehat\omega_i)\eta(\widehat\omega),\quad \forall i \in \text{supp}(\widehat \omega)\label{eqn:cnstr1}\\
        &\quad\quad |\omega_i|<\eta(\widehat \omega),\quad\quad\quad\quad \forall i \in \overline{\text{supp}}(\widehat \omega)\label{eqn:cnstr2}\\
        &\quad\quad\| \Upsilon_0\widetilde z - \Upsilon_0\Phi\omega\|_2 \leq \mu\label{eqn:cnstr3}
    \end{align}
    \label{eqn:optprob_ver}
\end{subequations}
where $\eta(\omega) \doteq \min\limits_{i\in\text{supp}(\omega)} |\omega_i|$.

\begin{theorem}[maximum sparsity recovery]
    \label{thm:cn}
    Let $\bar\omega$ and $\omega^v$ be the solutions of problems \eqref{eqn:opt_prob} and \eqref{eqn:optprob_ver}, respectively. Let $\widehat \omega$ be the parameter vector identified solving problem \eqref{eqn:optprob1}, ${M} \doteq \| \widehat\omega\|_0$ the cardinality of its support, and~${\Upsilon_0 = \Upsilon(\Xi)}$. Let~${\widetilde z = [\widetilde z_1,\dots, \widetilde z_T]^\top}$ the vector of noise corrupted observation collected from \eqref{eqn:system.bar}. Define $\kappa_e \doteq \|\widehat\omega\|_0-\|\bar\omega\|_0$ and the set $$\begin{aligned}
        &\lambda \doteq \\&\left\{i\!:\! |\omega_i^v|\!>\!\frac{\|e^{\star}_{\Xi,\Phi}(\widetilde z,\omega^v)\|_{(\Upsilon_0,1)}\!+\!\|e^{\star}_{\Xi,\Phi}(\widetilde z,\omega^v)\|_{(\Upsilon_0,2M)}}{\sigma_{min,2M}^2(\Upsilon_0)}\!\right\}\!.
    \end{aligned}$$
    Assume that the constraint \eqref{eqn:cnstr3} is active, i.e.
    \begin{equation*}
        \| \Upsilon_0\widetilde z - \Upsilon_0\Phi\omega^v\|_2 = \mu
    \end{equation*}
    Then, 
    \begin{equation}
        \kappa_e \leq \bar\kappa_e \doteq \|\widehat\omega\|_0 - card(\lambda).
        \label{eqn:thm2.1.ss}
    \end{equation}
    Moreover, if 
    \begin{equation}
         \frac{\|e^{\star}_{\Xi,\Phi}(\widetilde z,\omega^v)\|_{(\Upsilon_0,1)} + \|e^{\star}_{\Xi,\Phi}(\widetilde z,\omega^v)\|_{(\Upsilon_0,2M)}}{\sigma_{min,2M}^2(\Upsilon_0 )}<\eta(\widehat\omega),
         \label{eqn:thm2.ineq.ss}
    \end{equation}
    then $\widehat\omega$ is maximally sparse ($\bar\kappa_e=0$), and
    \begin{equation}
        \mathrm{supp}(\widehat\omega) = \mathrm{supp}(\bar\omega).
        \label{eqn:thm2.res.ss}
    \end{equation}
\end{theorem}
The proof of Theorem \ref{thm:cn} is reported in Appendix \ref{app:cn}.
\subsection{Multi-step, nonlinear-in-parameters extension}
Consider now a more general model, described by the system in \eqref{eqn:system} with parameters $\bar \theta$, $n_x=n_z=1$, and no assumptions on the shape of $f$ and $h$.
A dictionary of basis function, $\varphi \in \mathbb R^{m}$, and a set of noise-corrupted data $\mathcal D = \{\mathbf{\widetilde u}_{0:T}, \mathbf{\widetilde z}_{0:T}\}$ collected from \eqref{eqn:system} are available. Here, the measurement model is described by
$$
\widetilde z_{k+1} = f(x_k,\widetilde u_k, \bar \theta)  + \Delta(x_k,\widetilde u_k) + \eta^v_k
$$
according to the measurement model derived in Appendix \ref{app:measurement_model}.
\begin{problem}[multi-step sparsity problem] \label{prb:B}
    Defining the following parametrized, multi-step system
\begin{equation}
\begin{aligned}
{\mathcal{M}}: \widehat x_{k+1}&=\widehat F(\widehat 
 x_k, \widetilde u_k,\theta,\omega) \\&= f(\widehat  x_k,  \widetilde u_k, \theta) + \sum_{i=1}^m \omega_i\varphi_i(\widehat  x_k, \widetilde u_k),\\
\widehat z_k &= h(\widehat x_k).
\end{aligned}
\label{eqn:Fhat}
\end{equation}
the goal is to identify the coefficient vector $\widehat \theta, \widehat \omega$ from the data set $\mathcal D$, such that:
\begin{enumerate}[i)]
    \item  $\widehat \omega$ is sparse;
    \item $(\widehat \theta, \widehat \omega)$ are consistent with the dataset, i.e., $\|\widetilde z~-~\widehat z\|_2 \leq \mu$,
    with $\widehat z = [\widehat z_1,\dots, \widehat z_T]^\top$. 
\end{enumerate}
\end{problem}
\begin{remark}[multi-step nonlinearity]
    In the multi-step case, iterating the system over a horizon of length~${T>1}$ leads to a highly nonlinear-in-parameters system even if $\widehat F $ is assumed to be linear in $\theta$ and $\omega$. Indeed, the output equation in \eqref{eqn:Fhat}, representing the $k$-th output prediction, can be written iterating the recursion along the horizon as 
\begin{equation}
     \widehat z_k = 
     h(\widehat F^k(\widehat x_{0}, \widetilde u_1,\dots,\widetilde u_{k-1},\theta,\omega))
\label{eqn:Fhat2}
\end{equation}
where $\widehat x_{0}$ is a given estimation (or measurement) of the system's initial condition {and $\widehat F^k(\widehat x_{0}, \widetilde u_1,\dots,\widetilde u_{k-1},\theta,\omega)$ denotes the $k$-th iteration of \eqref{eqn:Fhat}}.
\end{remark}
In this case, a solution to the sparse identification Problem \ref{prb:B} is given by solving the multi-step optimization problem \eqref{eqn:opt_prob}, exploiting the sequence of multi-step predictions $\widehat{\mathbf{x}}_{0:T-1}$ as states.
As in the previous case, we aim to solve the following relaxed version,
\begin{equation}
    \begin{array}{cl}
        (\widehat \theta, \widehat \omega) = & \arg \min\limits_{\theta, \omega} \|\omega\|_1\\
        & \text{s.t. } \|\widetilde z - \widehat z\|_2 \leq \mu.
    \end{array}
    \label{eqn:optprob1_ms}
\end{equation}

In the following, we propose the extension of Theorems~\ref{thm:grib_var} and \ref{thm:cn} to the multi-step case, under Assumptions \ref{ass:noise} and~\ref{ass:feasibleprob}. 
{Note that, from a conceptual point of view, this extension is not trivial, since the parameters to estimate appear as arguments of highly nonlinear functions, resulting from the iterative process needed to compute the multi-step prediction. From a technical point of view, the extension is relatively simple, since carried out using standard Jacobian linearization.}

\begin{theorem}[local equivalent support conditions] \label{thm:grib_var_ms}
Let~$\widehat \omega \in \mathbb R^m$ be a local solution of the optimization problem \eqref{eqn:optprob1_ms} and ${M} \doteq \| \widehat\omega\|_0$ the cardinality of its support. Let $\omega'$ be any other representation in the local neighborhood of $\widehat \omega$. 
Define the Jacobian matrices
\begin{subequations}
\begin{align}
    \mathcal{J}_\theta &\doteq \frac{\partial\widehat F(\mathbf{\widehat x}_{0:T-1}, \mathbf{\widetilde u}_{0:T-1}, \widehat \theta, \widehat \omega)}{\partial    \theta} \in \mathbb R^{T,n_\theta},\\
    \mathcal{J}_\omega &\doteq \frac{\partial\widehat F(\mathbf{\widehat x}_{0:T-1}, \mathbf{\widetilde u}_{0:T-1}, \widehat \theta, \widehat \omega)}{\partial \omega} \in \mathbb R^{T,m},
\end{align}
\label{eqn:jacobians}
\end{subequations}
and $\Upsilon_\ell = \Upsilon(\mathcal{J}_\theta)$. Moreover, define
\begin{equation}
\widetilde z_\ell = \widetilde z - \widehat F(\mathbf{\widehat x}_{0:T-1}, \mathbf{\widetilde u}_{0:T-1}, \widehat \theta, \widehat\omega) - \mathcal{J}_\theta\widehat\theta - \mathcal{J}_\omega\widehat\omega.
\label{eqn:measvect_lin}
\end{equation}
If the following inequalities hold
    \begin{subequations}
    \begin{align}
    \|e^{\star}_{\mathcal{J}_\theta,\mathcal{J}_\omega}(\widetilde z_\ell,\omega')\|_2 &\leq \|e^{\star}_{\mathcal{J}_\theta,\mathcal{J}_\omega}(\widetilde z_\ell,\widehat \omega)\|_2,\\
        \|\omega'\|_0 &\leq M,
    \end{align}
    \end{subequations}
    and
    \begin{equation}
    \begin{aligned}
    \|e^{\star}_{\mathcal{J}_\theta,\mathcal{J}_\omega}(\widetilde z_\ell,\widehat\omega)\|_{(\Upsilon_\ell,1)} + \|&e^{\star}_{\mathcal{J}_\theta,\mathcal{J}_\omega}(\widetilde z_\ell,\widehat\omega)\|_{(\Upsilon_\ell,2M)}\\ &< \sigma_{min,2M}^2 (\Upsilon_\ell)\eta(\widehat \omega),
        \end{aligned}
    \end{equation}
    where $\eta(\omega) \doteq \min\limits_{i\in\text{supp}(\omega)} |\omega_i|$,
    then $\widehat\omega$ and $\omega'$ have the same support,
    $$\mathrm{supp}(\widehat\omega) = \mathrm{supp}(\omega'),$$ and the same sign, $$\mathrm{sign}(\widehat\omega_i) = \mathrm{sign}(\omega'_i), \quad \forall i\in[1,m].$$
\end{theorem} 
\noindent 
\textit{Proof.}$\quad$ The proof is derived considering the linearization of $\widehat F$ around the solution of \eqref{eqn:optprob1_ms} $(\widehat \theta, \widehat \omega)$, i.e.,
\begin{equation}
    \begin{aligned}
        \widehat F(&\mathbf{\widehat x}_{0:T-1}, \mathbf{\widetilde u}_{0:T-1}, \theta, \omega) \simeq \\ &\widehat F(\mathbf{\widehat x}_{0:T-1}, \mathbf{\widetilde u}_{0:T-1}, \widehat \theta, \widehat\omega) + \mathcal{J}_\theta(\theta-\widehat\theta) + \mathcal{J}_\omega(\omega-\widehat\omega),
    \end{aligned}
\end{equation}
and applying the proof of Theorem \ref{thm:grib_var}, reported in Appendix \ref{app:grib_var}, and Theorem 1, Corollary 1 in \cite{gribonval2006simple}, considering  $\Upsilon_\ell\widetilde z_\ell = \Upsilon_\ell\mathcal{J}_\omega\widehat\omega+e^{\star}_{\mathcal{J}_\theta,\mathcal{J}_\omega}(\widetilde z_\ell,\widehat\omega)$, yielding the proposed local result for the multi-step framework. 
\hfill $\blacksquare$

Let $\bar\omega_\ell$ be the maximally sparse vector in the local approximation around $(\widehat \theta, \widehat \omega)$, i.e., the solution of 
\begin{equation}
    \begin{array}{cl}
        \bar \omega_\ell \!=\!&\arg\min\limits_{\theta,\omega} \|\omega\|_0\\
        & \mathrm{{s.t.}}\; \|\widetilde z\!-\!\mathcal{J}_\theta\theta\!-\!\mathcal{J}_\omega \omega\|_2\!\leq\! \mu.
    \end{array}
    \label{eqn:opt_prob_lin}
\end{equation}
Hence, define the vector $\omega^v_\ell$ as the solution of the following optimization problem, i.e.,
\begin{subequations}
    \begin{align}
        \omega^v_\ell=&\arg \min\limits_{ \omega \in\mathbb R^m} \|\omega\|_1\\
        & \text{s.t. } \,\,\,\text{sign}(\widehat\omega_i)\omega_i \geq \eta(\widehat\omega),\quad \forall i \in \text{supp}(\widehat \omega)\label{eqn:cnstr1ms_lin}\\
        &\quad\quad |\omega_i|<\eta(\widehat \omega),\quad\quad\quad\quad \forall i \in \overline{\text{supp}}(\widehat \omega)\label{eqn:cnstr2ms_lin}\\
        &\quad\quad\|\Upsilon_\ell\widetilde z_\ell - \Upsilon_\ell\mathcal{J}_\omega\widehat\omega\|_2 \leq\mu.\label{eqn:cnstr3ms_lin}
    \end{align}
    \label{eqn:optprob_ver_ms_local}
\end{subequations} 
with $\eta(\omega) \doteq \min\limits_{i\in\text{supp}(\omega)} |\omega_i|$.
We formally define the conditions under which a vector is locally maximally sparse with the following theorem. Specifically, it allows us to check if an estimation of the parameters $(\widehat \theta,\widehat\omega)$, obtained as the solution to the optimization problem \eqref{eqn:optprob1_ms}, is the maximally sparse solution within the local neighborhood around the estimate, defined by the linearization.
\begin{theorem}[maximum sparsity local recovery]
    \label{thm:cn_ms}
    Let $\bar\omega_\ell$ and $\omega^v_\ell$ be the solutions of problems \eqref{eqn:opt_prob_lin} and \eqref{eqn:optprob_ver_ms_local}, respectively. Let $\widehat \omega$ be the parameter vector identified solving problem \eqref{eqn:optprob1_ms}, ${M} \doteq \| \widehat\omega\|_0$ the cardinality of its support, $\mathcal{J}_\theta$, $\mathcal{J}_\omega$ the jacobian matrices of $\widehat F$ with respect to $\theta$ and $\omega$ \eqref{eqn:jacobians}, $\widetilde z_\ell$ defined as in \eqref{eqn:measvect_lin}, and~$\Upsilon_\ell = \Upsilon(\mathcal{J}_\theta)$. Define $\kappa_e \doteq \|\widehat\omega\|_0-\|\bar\omega_\ell\|_0$ and the set 
    $$
    \begin{aligned}
    \lambda \doteq \Bigg\{&i\!:\! |\omega_{\ell,i}^v|\!>\\&\!\frac{\|e^{\star}_{\mathcal{J}_\theta, \mathcal{J}_\omega}\!(\widetilde z_\ell,\omega^v_\ell)\|_{(\Upsilon_\ell,1)}\!+\!\|e^{\star}_{\mathcal{J}_\theta, \mathcal{J}_\omega}\!(\widetilde z_\ell,\omega^v_\ell)\|_{(\Upsilon_\ell,2M)}}{\sigma_{min,2M}^2(\Upsilon_\ell)}\!\Bigg\}\!.
    \end{aligned}$$
    Assume that the constraint \eqref{eqn:cnstr3ms_lin} is active, i.e.
    \begin{equation*}
        \| \Upsilon_\ell\widetilde z - \Upsilon_\ell\Phi\omega^v\|_2 = \mu
    \end{equation*}
    Then, 
    \begin{equation}
        \kappa_e \leq \bar\kappa_e \doteq \|\widehat\omega\|_0 - card(\lambda).
        \label{eqn:thm2.1}
    \end{equation}
    Moreover, if 
    \begin{equation}
         \!\frac{\|e^{\star}_{\mathcal{J}_\theta, \mathcal{J}_\omega}(\widetilde z_\ell,\omega^v_\ell)\|_{(\Upsilon_\ell,1)}\!+\!\|e^{\star}_{\mathcal{J}_\theta, \mathcal{J}_\omega}(\widetilde z_\ell,\omega^v_\ell)\|_{(\Upsilon_\ell,2M)}}{\sigma_{min,2M}^2(\Upsilon_\ell)}\!\!<\eta(\widehat\omega),
         \label{eqn:thm2.ineq}
    \end{equation}
    then $\widehat\omega$ is locally maximally sparse ($\bar\kappa_e=0$), and
    \begin{equation}
        \mathrm{supp}(\widehat\omega) = \mathrm{supp}(\bar\omega_\ell).
        \label{eqn:thm2.res}
    \end{equation}
\end{theorem}
\textit{Proof.}$\quad$ The proof can be obtained by applying the same reasoning of the proof of Theorem \eqref{thm:cn}, reported in Appendix \ref{app:cn}, while referring to the proof of \eqref{thm:grib_var_ms}.
\hfill $\blacksquare$

\begin{remark}[Relation with the original problem]
    Although in this section the optimization problems are presented in the constrained form, as in \eqref{eqn:optprob1} and \eqref{eqn:optprob1_ms}, as discussed in \cite{gribonval2006simple} it is always possible to re-cast them in the unconstrained form \eqref{eqn:optprob.bb} considered within the proposed framework, by solving the corresponding Lagrangian problem for appropriate Lagrangian multipliers.
\end{remark}}

\subsection{Optimality of the physical parameters}\label{subsec:opt_err}
In this conclusive section, {we show that there exist situations where the physical parameters $\theta$, identified from the dataset $\mathcal{D}$ exploiting black box augmentation and maximum sparsity conditions, are \textit{provably better,} in the worst-case sense,  than those obtained with standard estimations with no sparsity guarantees.
In particular, we} assume that given a dictionary of basis functions $\varphi$ there exists a maximally sparse coefficient vector $\bar \omega$, solution of \eqref{eqn:opt_prob}, such that the unknown term $\Delta(x_k, u_k)$ of a system to identify \eqref{eqn:system} with true physical parameters $\bar \theta$ can be parameterized as
\begin{equation}
\Delta(x_k, u_k) = \sum_{i=1}^m \bar \omega_i \varphi_i(x_k, u_k).
\label{eqn:DeltaParam}
\end{equation}
It follows that the overall system has true parameters $\bar \theta$ and~$\bar \omega$.
Then, let us introduce the following definitions of feasible parameter sets on both the parameters $\theta$ and $\omega$.
\begin{definition}[parameters sets]
    Given a set of noise-corrupted data $\mathcal{D}=\{\mathbf{{\widetilde u}}_{0:T}, \mathbf{{\widetilde z}}_{0:T}\}$ satisfying Assumption \eqref{ass:noise}, a sequence of states (predicted or measured)~$\mathbf{x}_{0:T-1}$, and the bound on the noise $\mu$, the Feasible Parameter Set is defined as
    $$
    \begin{aligned}
    \mathrm{FPS} \doteq \Big\{\theta \in \mathbb R^{n_\theta}&, \omega \in \mathbb{R}^m: \\&\|\widetilde z - \widehat F(\mathbf{x}_{0:T-1}, \mathbf{\widetilde u}_{0,T-1},\theta,\omega)\|_2 \leq \mu\Big\}.
    \end{aligned}
    $$
    Moreover, the Supported Feasible Parameter Set is defined as
    $$\begin{aligned}
    \mathrm{SFPS} \doteq \Big\{\theta \in \mathbb R^{n_\theta}&, \omega \in \mathbb{R}^m: \\&\|\widetilde z - \widehat F(\mathbf{x}_{0:T-1}, \mathbf{\widetilde u}_{0,T-1},\theta,\omega)\|_2 \leq \mu, 
    \\&\mathrm{supp}(\omega)=\mathrm{supp}(\bar\omega) \Big\}\end{aligned}$$
    where $\bar \omega$ is the maximally sparse coefficients vector.
\end{definition}
According to these definitions, FPS and SPFS are the sets of all parameter vectors consistent with the dataset $\mathcal{D}$. Moreover, SPFS is the smallest set guaranteed to contain $(\bar \theta, \bar \omega)$:
$$(\bar \theta, \bar \omega) \in \mathrm{SFPS} \subseteq \mathrm{FPS}.$$

We now formally define the parametric error $e_\theta(\theta)$, i.e., a measure of the distance with respect to the true (physical) parameters vector: 
\[e_\theta(\theta) =\|\bar \theta - \theta \|_2\]
A tight bound on $e_\theta(\theta)$ is given by the worst-case parametric error, i.e., the error between the current estimate $\widehat\theta$ and the $\theta\in\text{SFPS}$ at a maximum distance from $\widehat\theta$, as stated in the following definition.
\begin{definition}[worst-case parametric error]
    The worst-case parametric error of an estimate $\widehat\theta$ is
    $$e^W_\theta(\widehat\theta) = \sup_{\theta\in \mathrm{SFPS}} \|\theta - \widehat \theta\|_2.$$
\end{definition}
%
The following theorem shows that finding the ``correct sparsity" of the black model allows for obtaining an \textit{optimal physical estimate}, i.e., an estimate 
with a lower worst-case parametric error, compared to  ``standard" estimates without sparsity guarantees or obtained without a black model augmentation. 
\begin{theorem}[maximum sparsity optimality] \label{thm:optimal_estimate}
Consider a realization of $\Delta$ of the form of \eqref{eqn:DeltaParam}. Consider an estimation $(\widehat \theta,\widehat \omega)$, solution of \eqref{eqn:optprob1_ms}, in the region of attraction of {$(\bar \theta,\bar \omega)$}. If the conditions of Theorem \ref{thm:cn_ms} hold, then the physical parameters vector $\widehat \theta$, identified from the dataset $\mathcal{D}$, is an optimal physical estimate.
\end{theorem}
\textit{Proof.}$\quad$ Assume that the conditions of Theorem \ref{thm:cn_ms} hold for $(\widehat \theta,\widehat \omega)$. It follows that $\text{supp}(\widehat \omega) = \text{supp}(\bar \omega)$. Thus, $\widehat \omega$ is maximally sparse and $(\widehat \theta, \widehat \omega) \in$ SFPS. 
Since $\text{SFPS} \subseteq \text{FPS}$ it follows that $e^W_\theta(\widehat\theta) \leq e^W_\theta(\theta)$, by definition, for all $\theta \in \{\text{FPS} \setminus \text{SFPS} \}$,
showing that finding the "correct sparsity" allows to obtain an optimal estimate with better estimation accuracy (in a worst-case sense) with respect to a “standard” estimate, which is only guaranteed to belong to FPS.
Also, this is an immediate consequence considering \eqref{eqn:DeltaParam} and the result of Theorem \ref{thm:bounds}. From Theorem \ref{thm:bounds} it follows that the more the black-box model is able to compensate for the unknown term, the more $\widetilde \Delta$ is small, reducing the bound on the parametric error. Consequently, in the scenario discussed in this section, if $\widehat\omega=\bar\omega$, then $\widetilde\Delta=0$. \hfill $\blacksquare$
\section{Numerical examples}\label{sec:num_res}
In this section, two numerical examples are provided. First, the results of Theorem~\ref{thm:exp.grad} are shown through simulation in Section~\ref{sec:expl_grad_ex} for the identification of the parameters of a population dynamics model. Then, we illustrate the effectiveness of the proposed framework in Section~\ref{sec:CTSbenchmark} by showcasing its results on a cascade tank system identification benchmark \cite{schoukens2016cascaded},
while comparing it with other state-of-the-art
identification methods applied to the same benchmark \cite{svensson2017flexible,schoukens2016modeling,brunot2017continuous,relan2017unstructured,mattsson2018identification,mavkov2020integrated,dalla2021kernel}.

\subsection{Exploding gradient: logistic map example} \label{sec:expl_grad_ex}
The discrete-time logistic map \cite{logistic_map_1}
is a dynamical system that exhibits complex behavior, including chaos. It is defined by the following recurrence relation
\begin{equation}
    x_{k+1} = \theta x_k (1-x_k),
    \label{eqn:logisticmap}
\end{equation}
where $x_k \in \mathbb R$ is the state, typically restricted to the interval $[0,1]$, while $\theta \in \mathbb R$ is a parameter that controls the behavior of the system.
Hence, the behavior of the logistic map depends crucially on $\theta$. In particular, we have that $x_k$ converges to $0$ for $0\leq \theta \leq 1$, to a fixed point in for $1<\theta\leq3$, or oscillates between two fixed points in $[0,1]$ for~${3<\theta<\theta_c}$, with $\theta_c \approx 3.57$. On the other hand, 
$x_k$ exhibits chaotic behavior in $[0,1]$ for $\theta_c<\theta\leq4$, while it leaves this interval and diverges in finite-time for $\theta>4$, for almost all initial conditions. 

Let us consider a population dynamics model described by the logistic map \eqref{eqn:logisticmap},
where $x_k$ represents the population size at time step $k$, and $\theta$ is the parameter representing the growth rate. We want to identify the value of $\theta$ that leads to a given, stable, nominal population evolution $\{\widetilde x_k\}_{k=1}^T$. 
Clearly, knowing the parameter values that lead to instability is beneficial, as this enables to impose directly parameter constraints.
For instance, we could set a constraint such as $0\le\theta\le 4$ to ensure system stability. However, in most cases, only the values of the state at which the system does not show instability are known.
In this case, {relying on the results in Section~\ref{sec:pbconstraints}}, it is possible to use a barrier function that bounds the predicted population size away from infinity in order to generate predicted trajectories that are not finite-time unstable thus ensuring non-exploding gradients and allowing the identification of $\theta$.  In particular, an exponential barrier function of the form 
\begin{equation}
    p(\widehat{x}_k) \doteq e^{\alpha(\hat{x}_k - 1)} + e^{\alpha(-\hat{x}_k)},
    \label{eqn:bf_exp}
\end{equation}
with $\alpha\in\mathbb R$ a sharpness parameter, can be used to bound the predicted trajectories to the interval $[0,1]$.

Let us now consider the specific case with true parameters $\bar \theta = 3.5$. We initialize the identification algorithm with $\widehat \theta_0 = 3.9$, aiming to identify the true value. This is a ``critical" area for the parameters, very close to the values of instability ($\theta>4$), and in the region where the system exhibits chaotic behavior. Therefore, it is likely that the gradient, when no barrier functions {were} used, will update the estimated parameter in the direction of a possible local minimum near the values of $\theta>4$, causing gradient explosion. 
\begin{figure}[!tb]
    \centering
    \includegraphics[width =  \columnwidth]{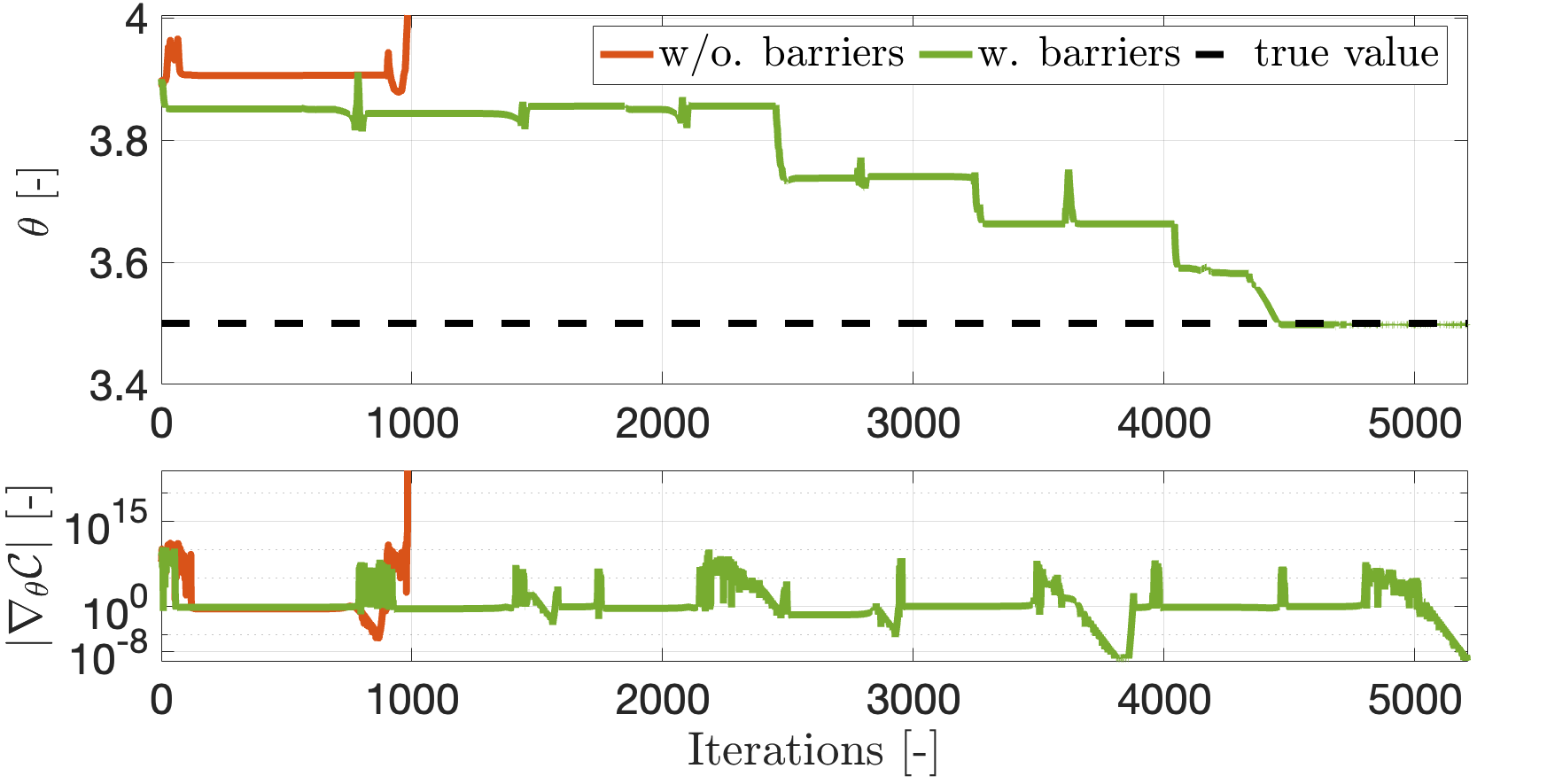}
    \caption{Effect of the use of barriers function to avoid exploding gradients.}
    \label{fig:bf_example}
\end{figure}
Fig. \ref{fig:bf_example}  shows the evolution of the estimated parameters along with the computed gradient when no barrier functions are used (orange line), and when the barrier function proposed in \eqref{eqn:bf_exp} with $\alpha = 75$ is adopted in the loss function (green line). It is possible to notice that when no barrier functions are used, the gradient explodes when the estimated parameter reaches the finite-time instability zone with $\theta>4$. On the other hand, the use of an exponential barrier function helps the gradient to steer the parameter in the chaotic area, avoiding values that cause instability, until reaching a neighborhood of the true values.
\subsection{Cascaded Tanks System benchmark} \label{sec:CTSbenchmark}
The cascaded tanks system (CTS) benchmark, thoroughly described in \cite{schoukens2016cascaded},
is used in this section to evaluate the proposed approach for system identification. 

The CTS controls fluid levels using two connected tanks with free outlets and a pump. An input signal controls a water pump that transfers water from a reservoir to the upper tank. Then, the water flows through a small opening to the lower tank and back into the reservoir.
Water overflow occurs when one tank is full. If the upper tank overflows, some water flows into the lower tank, while the rest leaves the system.
An approximate nonlinear, continuous, state-space model of the CTS is derived in \cite{schoukens2016cascaded} exploiting Bernoulli’s principle and conservation of mass, i.e.,
\begin{equation}
\begin{aligned}
    \dot x_1(t) &= -k_1 \sqrt{x_1(t)} + k_4 u(t) + v_1(t),\\
    \dot x_2(t) &= k_2 \sqrt{x_1(t)} - k_3\sqrt{x_2(t)} + v_2(t),\\
    z(t) &= x_2(t) + \eta^z(t),
    \label{eqn:CTS}
\end{aligned}
\end{equation}
where $u(t) \in \mathbb R$ is the input signal, $x_1(t)\in \mathbb R$ and $x_2(t)\in \mathbb R$ are the states of the system, $v_1(t)\in \mathbb R$, $ v_2(t)\in \mathbb R$ and $\eta^z(t)\in \mathbb R$ are additive noise sources. The system is characterized by four unknown physical constants, $k_1$, $k_2$, $k_3$, and $k_4$, which depend on the properties of the system and need to be estimated. It is important to note that the initial values of the system states are also unknown when measurements start, and thus need to be also estimated. This unknown state is the same for both the estimation and the validation data record. Moreover, since this model ignores the water overflow effect, unmodelled dynamics are present in the physical dynamics in \eqref{eqn:CTS}.
The training and the validation datasets consist of $T = 1024$ input and output samples and the sampling time is $T_s = 4 s$. 

The main objective is to accurately estimate the dynamics of the system using only the available training data. 
For the identification, the following discretized physical model is considered
\begin{equation}
\begin{aligned}
    x_{1,k+1} &= x_{1,k} + T_s\left(-k_1 \sqrt{x_{1,k}} + k_4 u_k + v_{1,k}\right),\\
    x_{2,k+1} &= x_{2,k} + T_s\left( k_2 \sqrt{x_{1,k}} - k_3\sqrt{x_{2,k}} + v_{2,k}\right),\\
    z_k &= x_{2,k} + \eta^z_k,
\end{aligned}
\label{eqn:tctmodel}
\end{equation}
and augmented, according to the model \eqref{eqn:bb_extension_v2}, with a black-box term consisting of sigmoids, softplus, hyperbolic, and trigonometric basis functions.
The effectiveness of the estimation algorithm is then evaluated on the validation dataset. As suggested in \cite{schoukens2016cascaded}, the Root Mean Square Error (RMSE) between the measured output $z$ and the predicted output $\widehat z$, i.e.,
\begin{equation}
    e_{RMS} = \sqrt{\frac1T\sum_{k=1}^{T}\|z_k-\widehat z_k\|_2^2}
\end{equation}
is used as a performance metric.
Being the proposed approach based on multi-step identification, $\widehat z$ is obtained through simulation, thus only the input $u$ is used to obtain the predicted output $\widehat z$.
A regularized local loss of the form \eqref{eqn:reg_term} is used to define the cost function $\mathcal{C}_T$, to ensure that the black-box component remains minimal if the physical model \eqref{eqn:tctmodel} adequately describes the system. In this example, we selected $\mathcal{L}_k = \frac 1T \|e_k\|_2^2$, $\rho = 0.1$, and~$\nu = 100$. The estimated parameters and initial conditions are initialized as $\widehat k_{1,0} = \widehat k_{2,0} = \widehat k_{3,0} = \widehat k_{4,0} = 0.05$, $\widehat x_{1,0}~=~\widehat x_{2,0}~=~z_0$.
After the optimization, the vector of black-box parameters has been further regularized. 
Specifically, each element of $\Omega$ is set to $0$ if and only if $|\Omega_{ij}|\leq10^{-4}$.

\begin{figure}[!tb]
    \centering
    \includegraphics[width = \columnwidth]{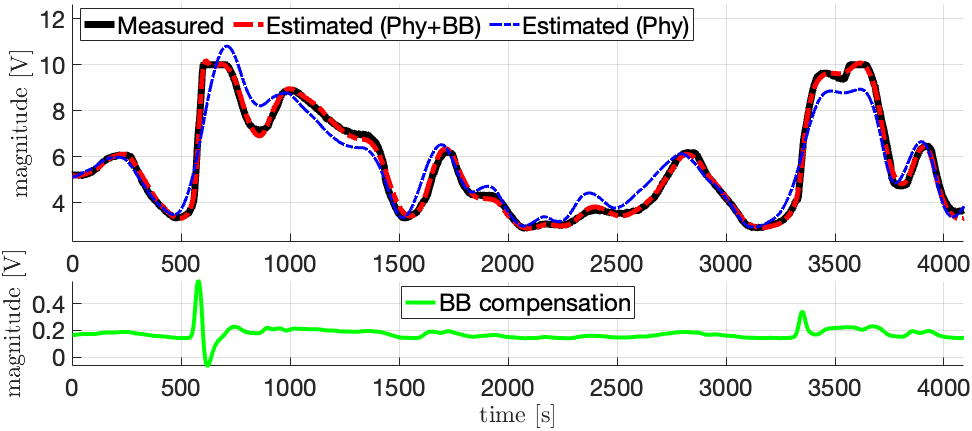}
    \caption{Training data (Simulation).}
    \label{fig:training}
\end{figure}

\begin{figure}[!tb]
    \centering
    \includegraphics[width = \columnwidth]{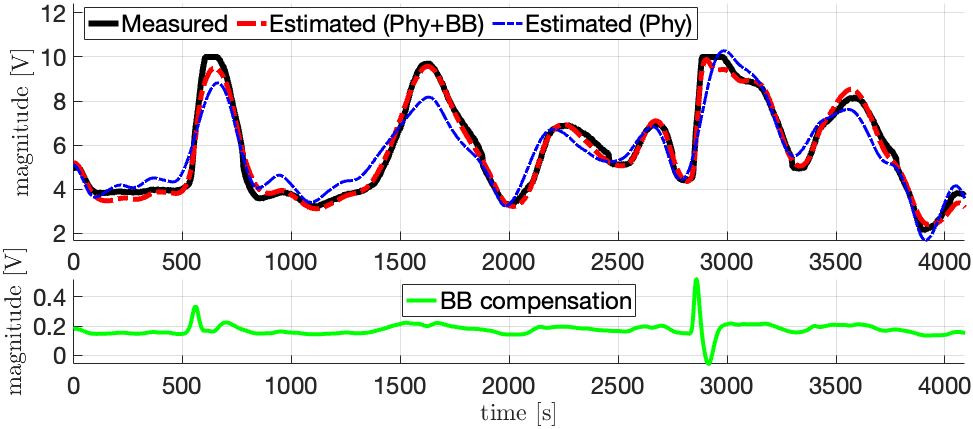}
    \caption{Validation data (Simulation).}
    \label{fig:test}
\end{figure}

The measured output and simulated model output in both the training and validation datasets are reported in Fig. \ref{fig:training} and Fig. \ref{fig:test}, along with the effect of the black-box compensation.
Moreover, a comparison is made between the predictions obtained using only the physical model (blue line) and those obtained with the black-box compensation (red line), showing an effective compensation of the unmodelled dynamics introduced by the overflow effect. 
The simulation shows that relying solely on the physical model may not achieve satisfactory results due to the presence of unmodeled overflow dynamics while incorporating a black-box augmented model proves to be beneficial in improving performance by accounting for these unmodeled terms.
In particular, the estimated model derived exclusively from the physical equations yields an RMSE index of $0.601$V for the training dataset and $0.668$V for the validation dataset. The model incorporating black-box compensation exhibits a significantly reduced RMSE index of $0.134$ V for the training dataset and $0.259$ V for the validation dataset.
%

Then, similarly to what is done in \cite{mavkov2020integrated}, we compare our approach and other state-of-the-art identification methods applied to the same benchmark. The resulting RMSE values of these algorithms for the validation data are given in Tab. \ref{tab:comp_tanks}. 
\begin{table}[!tb]
    \centering
    \caption{Comparison for different identification methods.}
    \begin{tabular}{l c}
    \hline
        Method & $e_{RMS}$\\
    \hline
         Svensson et al. \cite{svensson2017flexible} & $0.45$\\
         Volt.FB (Schoukens et al.) \cite{schoukens2016modeling} & $0.39$\\
         NOMAD (Brunot et al.) \cite{brunot2017continuous} & $0.37$\\
         PNLSS-I (Relan et al.) \cite{relan2017unstructured} & $0.45$\\
         NLSS2 (Relan et al.) \cite{relan2017unstructured}& $0.34$\\
         PWARX (Mattsson et al.) \cite{mattsson2018identification} & $0.35$\\
         INN (Mavkov et al.) \cite{mavkov2020integrated}  & $0.41$\\
         SED-MPK (Della Libera et al.) \cite{dalla2021kernel} & 0.48\\
         \hline
         \textbf{Proposed approach - Training Data} & $0.13$\\
         \textbf{Proposed approach - Validation Data} & $0.26$\\
    \hline
    \end{tabular}
    \label{tab:comp_tanks}
\end{table}
Notably, the proposed approach demonstrates superior performance, with the lowest performance metric values of $0.13$ for the training data and $0.26$ for the validation data, highlighting its effectiveness and robustness compared to existing methods. Moreover, it is important to note that, differently from other methods, we also estimate physical parameters and initial conditions. Specifically, we obtained $\widehat k_1=0.0764$, $\widehat k_2=0.0268$, $\widehat k_3=0.0415$, $\widehat k_4=0.0386$, $\widehat x_{1,0}=3.52$ V, and $\widehat x_{2,0}=5.19$ V.
\section{Conclusions and future works}\label{sec:concl}
In this paper we introduced a unified framework for identifying interpretable nonlinear dynamical models while preserving physical properties. It represents a substantial advancement in the use of combined off-white and sparse black models for system identification, allowing the compensation of unmodeled dynamics in physical models. Theoretical analysis established bounds on parameter estimation errors and identified conditions for gradient stability and sparsity recovery. The effectiveness of the approach has been demonstrated on a nonlinear system identification benchmark, resulting in significant improvements in model accuracy.

Building on the results of this study, future research should target several key areas to further enhance the nonlinear system identification framework. For instance, the versatility of the cost function allows greater adaptability of the optimization problem to specific scenarios, such as missing measurements, cumulative measurements, variable noise, and weighted measurements. Additionally, the proposed approach can be extended using various techniques to model the black-box term $\delta$, such as neural networks or kernel methods. Moreover, it can be further enhanced by incorporating a regressor of the states.
\begin{ack}                               
The authors thank Prof. Giuseppe Calafiore for his constructive feedback and insightful comments, which have helped to improve the quality of the paper.
\end{ack}      

\appendix
\section*{Appendix}
\section{Measurement model} \label{app:measurement_model}
Given a sequence of $T$ collected inputs $\mathbf{{\widetilde u}}_{0:T}$ applied to $\mathcal{S}$, and the corresponding $T$ collected observations $\mathbf{{\widetilde z}}_{0:T}$ with~${\widetilde u_k = u_k + \eta^u_k}$, and~${\widetilde z_k = z_k + \eta^z_k}$, we have
\begin{equation}
\begin{aligned}
    \widetilde z_{k+1} &= h(f\left({x}_{k}, {u}_{k}, \theta\right) + \Delta(x_{k},{u}_{k})) + \eta^z_{k+1},
    \\
    &\doteq \bar h\left({x}_{k}, {u}_{k}, \theta\right) + \eta^z_{k+1}
    \end{aligned}
\end{equation}
Substituting ${u_k = \widetilde u_k - \eta^u_k}$ we obtain
\begin{equation}
   \widetilde z_{k+1} =  \bar h\left({x}_{k}, \widetilde u_k - \eta^u_k, \theta\right) + \eta^z_{k+1}.
\end{equation}
Thus, using the mean value theorem we have 
$$
\begin{aligned}
\bar h\left({x}_{k}, \widetilde u_k, \theta\right)-\bar h\left({x}_{k}, \widetilde{u}_{k} - \eta^u_k, \theta\right) = \frac{\partial \bar h}{\partial \eta_{k}^{u}}\, \eta_{k}^{u}.
\end{aligned}
$$
so that, defining $\eta^v_k = \frac{\partial \bar h}{\partial \eta_{k}^{u}}(\eta_{k}^{u}) + \eta^z_k$, we can write the direct relationship between measured input and measured output as
\begin{equation}
\begin{aligned}
    \widetilde z_{k+1} &= h(f\left({x}_{k},\widetilde{u}_{k}, \theta\right) + \Delta(x_{k},\widetilde{u}_{k})) + \eta^v_k\\&= \bar h\left({x}_{k}, \widetilde u_k, \theta\right) + \eta^v_k.
    \label{eqn:meas_model}
\end{aligned}
\end{equation}
The relationship in \eqref{eqn:meas_model} represents the measurement model, that accounts for both measurement and process noise in the noise term $\eta^v_k$. 

\section{Proofs of Proposition \ref{prop1} and Proposition \ref{prop2}}
\subsection{Proof of Proposition \ref{prop1}}
\label{app:prop1}
Considering \eqref{eqn:cost}, the gradient with respect to $\theta$ can be computed as
\begin{equation*}
    \nabla_{{ \theta}}\mathcal{C}_{T} \doteq \frac{\d\mathcal{C}_T}{\d{\theta}} = \frac{\d}{\d{\theta}} \left( \sum_{k=1}^{T} \mathcal{L}_k \right) = \sum_{k=0}^{T} \frac{\d\mathcal{L}_k}{\d{\theta}}.
\end{equation*}
It follows that
\begin{equation}
    \nabla_{{\theta}}\mathcal{C}_{k} = \nabla_{{\theta}}\mathcal{C}_{k-1} + \frac{\d\mathcal{L}_k}{\d{\theta}},
    \label{eqn:cost_upd}
\end{equation}
where 
$
\nabla_{{\theta}}\mathcal{C}_{k} \doteq \sum_{\tau=0}^{k} \frac{\d\mathcal{L}_\tau}{\d{\theta}}
$ and $ \nabla_{{\theta}}\mathcal{C}_{0} = \mathbf{0}_{n_\theta} $.

Moreover, exploiting the chain rule of differentiation we have the following relation
\begin{equation}
    \begin{aligned}
        \frac{\d\mathcal{L}_k}{\d{ \theta}}^\top &=  
        {\frac{\partial \mathcal{L}_k}{\partial {\theta}}}^\top+
        \frac{\partial \mathcal{L}_k}{\partial e_k}^\top 
        \frac{\partial e_k}{\partial z_k}
        \frac{\partial z_k}{\partial {x_k}}
        \frac{\d {x_k}}{\d {\theta}}\\
    \end{aligned}
    \label{eqn:gradient_par}
\end{equation}
where the last term is defined as
\begin{equation}
    \frac{\d {x_k}}{\d {\theta}} = \frac{\partial {x_k}}{\partial {\theta}} + \frac{\partial {x_k}}{\partial {x_{k-1}}}\frac{\d x_{k-1}}{\d {\theta}}.
    \label{eqn:influence}
\end{equation}
%
Introducing the recursive memory operator $\Lambda_{k}$ defined in \eqref{eqn:Lambdadef}, we can rewrite \eqref{eqn:influence} as follows
\begin{equation}
    \Lambda_{k} = 
    \frac{\partial {x_k}}{\partial {\theta}} + \frac{\partial {x_k}}{\partial {x_{k-1}}}
    \Lambda_{k-1} = 
    \mathcal{J}^{x\!/\!\theta}_k + \mathcal{J}^{x\!/\!x}_k
    \Lambda_{k-1},
    \label{eqn:influence_upd}
\end{equation}
which yields \eqref{eqn:GUlaw_mat}. Thus, \eqref{eqn:gradient_par} can be rewritten as
\begin{equation}
     \frac{\d\mathcal{L}_k}{\d{\theta}}^\top ={\nabla_{\theta}^\top \mathcal{L}_k} + 
        {\nabla_{e}^\top \mathcal{L}_k} 
        {\mathcal{J}^{e\!/\!z}_k}
        {\mathcal{J}^{z\!/\!x}_k}
        \Lambda^\top_k = \varrho^\top_k + \rho^\top_k\Lambda_k,
        \label{eqn:grad_upd}
\end{equation}
using \eqref{eqn:rho-var-def}.
Thus, \eqref{eqn:GUlaw_grad} is obtained by transposing and substituting \eqref{eqn:grad_upd} in \eqref{eqn:cost_upd}, concluding the proof.

\subsection{Proof of Proposition \ref{prop2}}
\label{sec-appendix1}
Following the same reasoning of Appendix \ref{app:prop1}, consider
\begin{equation}
    \nabla_{{x_0}}\mathcal{C}_{k} = \nabla_{{x_0}}\mathcal{C}_{k-1} + \frac{\d\mathcal{L}_k}{\d x_0}.
    \label{eqn:cost_upd_x0}
\end{equation}
Exploiting the chain rule we have
\begin{equation}
    \frac{\d\mathcal{L}_k}{\d{x}_{0}}^\top = 
    \frac{\partial \mathcal{L}_k}{\partial e_k}^\top 
    \frac{\partial e_k}{\partial z_k}
    \frac{\partial z_k}{\partial {x_k}}
    \frac{\d x_k}{\d {x}_{0}},
    \label{eqn:gradient_par_x0}
\end{equation}
where
\begin{equation}
    \frac{\d x_k}{\d { x}_{0}} = 
    \frac{\partial x_k}{\partial {x_{k-1}}}
    \frac{\d x_{k-1}}{\d {x}_{0}}.
    \label{eqn:influence_x0}
\end{equation}

Using \eqref{eqn:phi0def},
\eqref{eqn:influence_x0} can be rewritten as
\begin{equation}
    \Lambda_{0,k} = 
    \frac{\partial x_k}{\partial {x_{k-1}}}
    \Lambda_{0,k-1} = \mathcal{J}^{x\!/\!x}_k\Lambda_{0,k-1}.
    \label{eqn:influence_upd_x0}
\end{equation}
which yields \eqref{eqn:GUlaw_initcond_mat}. Then, \eqref{eqn:gradient_par_x0} can be rewritten as
\begin{equation}
    \begin{aligned}
        \frac{\d\mathcal{L}_k}{\d x_0}^\top = 
        {\nabla_{e}^\top \mathcal{L}_k} 
        {\mathcal{J}^{e\!/\!z}_k} 
        {\mathcal{J}^{z\!/\!x}_k} 
        \Lambda_{0,k} = \rho^\top_k{\Lambda_{0,k}},
    \end{aligned}
    \label{eqn:grad_upd_x0}
\end{equation}
using \eqref{eqn:rho-var-def}.
Thus, \eqref{eqn:GUlaw_initcond} is obtained by transposing and substituting \eqref{eqn:grad_upd_x0} in \eqref{eqn:cost_upd_x0}. 

\section{Proofs of Theorem \ref{thm:exp.grad} and Corollary \ref{cor:corollary.exp.grad}}
\subsection{Proof of Theorem \ref{thm:exp.grad}}\label{app:exp.grad}
We observe that \eqref{eqn:GUlaw_grad} is a linear, time-varying (LTV) state-space system with
with states $\nabla_{{\theta}} \mathcal C_{k} \in \mathbb R^{n_\theta}$, and
\begin{equation}
\begin{aligned}
A_k &= A \doteq I_{n_\theta} \in \mathbb R^{n_\theta,n_\theta},\quad B_k \doteq (\Lambda^\top_k\rho_k + \varrho_k) \in \mathbb R^{n_\theta},\\
C_k &= C \doteq I_{n_\theta} \in \mathbb R^{n_\theta,n_\theta},\quad D_k = D \doteq \mathbf{0}_{n_\theta} \in \mathbb R^{n_\theta},
\end{aligned}
\label{matrixLTV}
\end{equation}
with constant input $u_k = 1 \in \mathbb R, \forall k \in [0,T]$.
By analyzing the bounded-input, bounded-output (BIBO) stability properties of the LTV system defined by \eqref{matrixLTV}, we can study the exploding gradient phenomenon and obtain conditions under which the multi-step gradient remains bounded.
The input-output behavior of \eqref{eqn:GUlaw_grad} is
specified by the unit-pulse response
$$
G(k,j)=C_k\Phi(k,j+1)B_j,\quad k\geq j+1
$$
with $\Phi(k,j)$ the transition matrix defined for $k>j$ as
$$
\Phi(k,j) = \left\{
\begin{array}{ll}
     A_{k-1}A_{k-2}\dots A_j & k\geq j+1\\
     I & k=j .
\end{array}
\right.
$$
Stability results are characterized in terms of boundedness properties of $G(k, j)$.
From Theorem 27.2 in \cite{rugh1996linear} we have that the linear state equation \eqref{eqn:GUlaw_grad} is uniformly BIBO stable if and only if there exists a finite constant $p$ such that the unit-pulse response satisfies
$$
\sum_{i=j}^{k-1}\|G(k,i)\|\leq p
$$
for all $k,j$ with $k\geq j+1$.
Notice that for our system defined by \eqref{matrixLTV}, we have 
$$
\Phi(k,j) = I_{n_\theta},\quad \forall k,j,  
$$
and
$$
G(k,j)=B_j, \quad k\geq j+1,
$$
that yields \eqref{eqn:theo2cond}, concluding the proof.
\subsection{Proof of Corollary \ref{cor:corollary.exp.grad}} \label{app:corollary.exp.grad}
Considering \eqref{eqn:theo2cond} on the finite time interval $k \in [1, T]$, it is worth noting that the condition holds if $\Lambda_i$, $\rho_i$, $\varrho_i$ are bounded for all $i \in [j, k-1]$, for all possible $k \in [1, T]$ with $k\geq j+1$. 
The problem is hence reduced to understanding under which conditions $\Lambda_i$, $\rho_i$, $\varrho_i$ may not be bounded. Here, it follows from Lipschitz continuity that the functions $\rho_k$, $\varrho_k$ \eqref{eqn:rho-var-def} are always bounded since $\mathcal{L}_k(\cdot)$, $e_k$, $h(\cdot)$ are at least continuously differentiable. 
On the other hand, the evolution over time of $\Lambda_k$ is given by \eqref{eqn:GUlaw_mat}. Here, \eqref{eqn:GUlaw_mat} can be seen as a linear, time-varying state-space system 
with states $x_k = \text{vec}{(\Lambda_{k})} \in \mathbb R^{n_xn_\theta}$, input $u_k=\text{vec}{(\mathcal{J}^{x\!/\!\theta}_k)}\in \mathbb R^{n_xn_\theta}$, $A_k=I_{n_\theta} \otimes \mathcal{J}_k^{x\!/\!x}\in \mathbb R^{n_xn_\theta,n_xn_\theta}$. Here, notice that the stability of this system is directly related to the matrix $\mathcal{J}_k^{x\!/\!x}$, corresponding to the linearization of the nonlinear system under analysis around the predicted trajectory obtained with the current values of~$\widehat \theta$ and $\widehat x_0$. This links the condition \eqref{eqn:theo2cond} directly to the stability of the predicted trajectories of the system. Hence, for the gradient to be bounded, the predicted trajectory must not diverge to infinity within the selected finite prediction horizon interval.

\section{Proof of Theorem \ref{thm:bounds}}\label{app:bounds}
Consider first the case where the dynamical system obeys the state representation \eqref{eqn:system}, with true parameters $\bar \theta \in \mathbb R^{n_\theta}$ and $\Delta = \delta = 0$. We have observed that $\mathcal{C}_{T}$ is a function
of the noise sequences, implying that also its minimizer $\theta^{\star}$
is a function of them, i.e., $\theta^{\star}\equiv\theta^{\star}(\boldsymbol{\eta})$.
Therefore,
\begin{equation}
\theta^{\star}-\bar{\theta}=\theta^{\star}(\boldsymbol{\eta})-\theta^{\star}(\mathbf{0}).\label{eq:par_err1}
\end{equation}
According to the Mean Value Theorem, a $\breve{\boldsymbol{\eta}}$ exists,
such that
\begin{equation}
\begin{aligned} \theta^{\star}(\boldsymbol{\eta})-\theta^{\star}(\mathbf{0}) &= \left.\frac{\partial\theta^{\star}(\boldsymbol{\eta})}{\partial\boldsymbol{\eta}}\right|_{\boldsymbol{\eta}=\breve{\boldsymbol{\eta}}}\boldsymbol{\eta} =\frac{\partial\theta^{\star}(\breve{\boldsymbol{\eta}})}{\partial\boldsymbol{\eta}}\boldsymbol{\eta}\\&=\left[\frac{\partial\theta^{\star}(\breve{\boldsymbol{\eta}})}{\partial\boldsymbol{\eta}}\right]_{u}\boldsymbol{\eta}^{u}_{1:T}+\left[\frac{\partial\theta^{\star}(\breve{\boldsymbol{\eta}})}{\partial\boldsymbol{\eta}}\right]_{z}\boldsymbol{\eta}^{z}_{1:T}
\end{aligned}
\label{eq:par_err2}
\end{equation}
where $\left[\frac{\partial\theta^{\star}(\breve{\boldsymbol{\eta}})}{\partial\boldsymbol{\eta}}\right]_{u}$
and $\left[\frac{\partial\theta^{\star}(\breve{\boldsymbol{\eta}})}{\partial\boldsymbol{\eta}}\right]_{z}$
are the matrices containing the columns of $\frac{\partial\theta^{\star}(\breve{\boldsymbol{\eta}})}{\partial\boldsymbol{\eta}}$
corresponding to $\boldsymbol{\eta}^{u}_{0:T}$ and $\boldsymbol{\eta}^{z}_{0:T}$, respectively.

In the following, we make use of implicit differentiation techniques to compute $\frac{\partial\theta^{\star}}{\partial\boldsymbol{\eta}}$.
Consider that the minimizer $\theta^{\star}$ is a solution of the
equation
$
\frac{\partial\mathcal{C}_{T}(\boldsymbol{\eta},\theta)}{\partial\theta}=0.
$
It follows that
\[
\frac{\d}{\d\boldsymbol{\eta}}\left.\frac{\partial\mathcal{C}_{T}(\boldsymbol{\eta},\theta)}{\partial\theta}\right|_{\theta=\theta^{\star}}=0
\]
where $\frac{\d}{\d\boldsymbol{\eta}}$ denotes the Jacobian with respect
to $\boldsymbol{\eta}$, given by
\[
\frac{\d}{\d\boldsymbol{\eta}}\left.\frac{\partial\mathcal{C}_{T}(\boldsymbol{\eta},\theta)}{\partial\theta}\right|_{\theta=
\theta^{\star}}
\!=\!\underbrace{\frac{\partial^{2}\mathcal{C}_{T}(\boldsymbol{\eta},\theta^{\star})}{\partial\boldsymbol{\eta}\partial\theta}}_{G}+\underbrace{\frac{\partial^{2}\mathcal{C}_{T}(\boldsymbol{\eta},\theta^{\star})}{\partial^{2}\theta}}_{H}\frac{\partial\theta^{\star}}{\partial\boldsymbol{\eta}}.
\]
The Hessian $H$ is invertible by assumption. Hence,
\[
\frac{\partial\theta^{\star}}{\partial\boldsymbol{\eta}}=-H^{-1}G(\boldsymbol{\eta}).
\]
Thus, it follows from (\ref{eq:par_err1}) and
(\ref{eq:par_err2}), recalling the definitions of $M_{u}$ and $M_{z}$ in Lemma \ref{lemma:bounded}, that
\begin{equation}
\left\Vert \theta^{\star}-\bar{\theta}\right\Vert _{p}\leq M_{u}\left\Vert \boldsymbol{\eta}^{u}_{0:T}\right\Vert_p+M_{z}\left\Vert \boldsymbol{\eta}^{z}_{0:T}\right\Vert_p.
\label{eqn:midresult}
\end{equation}
Consider now $\Delta \neq 0$. We observe that $\mathcal{C}_{T}$ is also a function
of $\boldsymbol{\widetilde \Delta}$ for the case $\Delta\neq 0$, implying that its minimizer, $\theta^{\star}$,
is not only a function of the noise sequences but also a function of $\boldsymbol{\widetilde \Delta}$, i.e., $\theta^{\star}\equiv\theta^{\star}(\boldsymbol{\eta}, \boldsymbol{\widetilde \Delta})$.
Therefore we can write,
\begin{equation}
\begin{aligned}
\theta^{\star}-\bar{\theta}&=\theta^{\star}(\boldsymbol{\eta}, \boldsymbol{\widetilde \Delta})-\theta^{\star}(\mathbf{0}, \mathbf{0})\\
&=\theta^{\star}(\boldsymbol{\eta}, \boldsymbol{\widetilde \Delta})-\theta^{\star}(\mathbf{0}, \mathbf{0}) + \theta^{\star}(\mathbf{0}, \boldsymbol{\widetilde \Delta}) - \theta^{\star}(\mathbf{0}, \boldsymbol{\widetilde \Delta})\\
&= [\theta^{\star}(\boldsymbol{\eta}, \boldsymbol{\widetilde \Delta}) - \theta^{\star}(\mathbf{0}, \boldsymbol{\widetilde \Delta})] + [\theta^{\star}(\mathbf{0}, \boldsymbol{\widetilde \Delta}) -\theta^{\star}(\mathbf{0}, \mathbf{0})].
\end{aligned}\label{eq:par_err11}
\end{equation}
Applying the implicit function theorem \cite{krantz2002implicit} to the function 
$P = \left.\frac{\partial\mathcal{C}_{T}(\mathbf{0}, \widetilde\Delta,\theta)}{\partial\theta}\right|_{\theta=\theta^{\star}}$ 
it follows that~$\theta^\star(\mathbf{0}, \boldsymbol{\widetilde \Delta})$ is continuously differentiable with respect to $\boldsymbol{\widetilde \Delta}$. Thus, from Lipschitz continuity, we have that there exists a constant $M_\Delta<\infty$ such that
\begin{equation}
    \|\theta^\star(\mathbf{0}, \boldsymbol{\widetilde \Delta}) - \theta^\star(\mathbf{0}, \mathbf{0})\|_p \leq M_\Delta\|\boldsymbol{\widetilde \Delta}\|_p.
    \label{eqn:b1}
\end{equation}
Considering $\eqref{eqn:midresult}$ when $\Delta \neq 0$, the following relation is obtained for all $\boldsymbol{\widetilde \Delta}$,
\begin{equation}
    \|\theta^{\star}(\boldsymbol{\eta}, \boldsymbol{\widetilde \Delta}) - \theta^{\star}(\mathbf{0}, \boldsymbol{\widetilde \Delta})\|_p \leq M_{u}\left\Vert \boldsymbol{\eta}^{u}_{0:T}\right\Vert_p+M_{z}\left\Vert \boldsymbol{\eta}^{z}_{0:T}\right\Vert_p.
    \label{eqn:b2}
\end{equation}
Thus, from \eqref{eqn:b1} and \eqref{eqn:b2}, we have 
\begin{equation}
\begin{aligned}
    \|\theta^{\star}(\boldsymbol{\eta}, \boldsymbol{\widetilde \Delta}) &- \theta^{\star}(\mathbf{0}, \boldsymbol{\widetilde \Delta})\|_p + \|\theta^\star(\mathbf{0}, \boldsymbol{\widetilde \Delta}) - \theta^\star(\mathbf{0}, \mathbf{0})\|_p \\
    &\leq M_{u}\left\Vert \boldsymbol{\eta}^{u}_{0:T}\right\Vert_p+M_{z}\left\Vert \boldsymbol{\eta}^{z}_{0:T}\right\Vert_p + M_\Delta\|\boldsymbol{\widetilde \Delta}\|_p.
\end{aligned}
\label{eqn:bound_Delta}
\end{equation}
The statement of the theorem follows from \eqref{eq:par_err11} and \eqref{eqn:bound_Delta}, recalling the triangle inequality.

\section{Proofs of Theorem \ref{thm:grib_var} and Theorem \ref{thm:cn}}
\subsection{Proof of Theorem \ref{thm:grib_var}}\label{app:grib_var}
The optimization problem \eqref{eqn:optprob1} is a \textit{feasibility problem} with respect to $\theta$. Therefore, we can reformulate the problem by selecting the appropriate values of $\omega$ for which a suitable $\theta$ exists that satisfies the constraint.
Hence, we consider the following equivalent
formulation
\begin{equation*}
\begin{aligned}
    \widehat \omega = &\arg \min\limits_{\omega} \|\omega\|_{1}\\
    &\text{s.t. } \omega \in \left\{ \omega: \exists \theta \text{ s.t. } \|{\widetilde z} - \Xi\theta - \Phi\omega\|_2 \leq \mu \right\}.
\end{aligned}
\end{equation*}

Moreover, since there exists at least one $\theta \in \mathbb R^{n_\theta}$ for which the constraint is satisfied, let us further reformulate the problem by seeking the $\omega$ for which $\theta \in \mathbb R^{n_\theta}$ minimizes the error and satisfy the constraints, i.e.,
\begin{subequations}
\begin{align}
    \widehat \omega = &\arg \min\limits_{\omega} \|\omega\|_{1}\\
    & \text{s.t. } \omega \in \biggl\{ \omega: \!\min_{\theta \in \mathbb R^{n_\theta}} \|{\widetilde z} - \Xi\theta - \Phi\omega\|_2 \!\leq\! \mu \biggr\}\!.
    \label{eqn:optprob122bcons}
\end{align}
\label{eqn:optprob122b}
\end{subequations}

Here, the $\theta$ minimizing $\|{\widetilde z} - \Xi\theta - \Phi\omega\|_2$ is given the least squares optimal solution, i.e.
$$\theta^\star(\omega) = (\Xi^\top\Xi)^{-1}\Xi^\top({\widetilde z} -\Phi\omega) = \Xi^\dag({\widetilde z} -\Phi\omega),$$ 
that once substituted in the least square cost in \eqref{eqn:optprob122bcons} gives
$$
\begin{aligned}
g(\omega) &= \|\widetilde z - \Xi\Xi^\dag(\widetilde z-\Phi\omega) - \Phi\omega\|_2 \\
&= \|\widetilde z - \Xi\Xi^\dag \widetilde z + \Xi\Xi^\dag \Phi\omega - \Phi\omega\|_2 \\
&= \|(I_T - \Xi\Xi^\dag) \widetilde z - (I_T -\Xi\Xi^\dag)\Phi\omega\|_2 \\
&= \| \Upsilon_0(\widetilde z - \Phi\omega)\|_2 = \|e^\star_{\Xi,\Phi}(\widetilde z, \omega)\|_2
\end{aligned}
$$
allowing to write \eqref{eqn:optprob122b} as
\begin{equation*}
    \begin{aligned}
        \widehat \omega =\arg &\min\limits_{\omega}
        \quad \|\omega\|_{1} \\
        & \text{s.t. } \,\, \|\Upsilon_0(\widetilde z - \Phi\omega)\|_2\leq \mu,
    \end{aligned}
\end{equation*} 
for which the results of Theorem 1 and Corollary 1 in \cite{gribonval2006simple} can be directly applied considering $\Upsilon_0\widehat z = \Upsilon_0\Phi\widehat\omega+e^{\star}_{\Xi,\Phi}(\widetilde z,\widehat\omega)$, concluding the proof.

\subsection{Proof of Theorem \ref{thm:cn}}\label{app:cn}
Notice that, following the same reasoning in the proof of Theorem \ref{thm:grib_var}, the optimization problem \eqref{eqn:opt_prob} with respect to $\omega$ can be written as 
\begin{equation*}
    \begin{aligned}
        \bar \omega =\arg &\min\limits_{\omega \in \mathbb R^m}
        \quad \|\omega\|_{0} \\
        & \text{s.t. } \,\, \|\Upsilon_0\widetilde z - \Upsilon_0\Phi\omega\|_2\leq \mu.
    \end{aligned}
\end{equation*} 
Hence, by definition, $\bar\omega$ is the sparsest vector satisfying the constraint $\|\Upsilon_0\widetilde z - \Upsilon_0\Phi\omega\|_2\leq \mu$.

Similarly to what is done in \cite{novara2012sparse}, consider $\omega^v$, solution of \eqref{eqn:optprob_ver}. Since the constraint \eqref{eqn:cnstr3} is active, we have that $$
\|e^{\star}_{\Xi,\Phi}(\widetilde z,\bar\omega)\|_2 \leq \|e^{\star}_{\Xi,\Phi}(\widetilde z,\omega^v)\|_2 = \mu.
$$ 
Moreover, $\|\bar\omega\|_0\leq\|\omega^v\|_0$, since $\bar\omega$ is maximally sparse. Then, according to Theorem \ref{thm:grib_var}, the inequality 
\begin{equation*}
    \|\bar\omega-\omega^v\|_\infty \leq \frac{\|e^{\star}(\widetilde z,\omega^v)\|_{(\Upsilon_0,1)} + |e^{\star}(\widetilde z,\omega^v)|_{(\Upsilon_0,2M)}}{\sigma_{min,2M}^2(\Upsilon_0 )}
\end{equation*}
holds, implying that, for all $i \in [1,m]$,
$$|\bar\omega_i - \omega_i^v| \leq \frac{\|e^{\star}(\widetilde z,\omega^v)\|_{(\Upsilon_0,1)} + \|e^{\star}(\widetilde z,\omega^v)\|_{(\Upsilon_0,2M)}}{\sigma_{min,2M}^2(\Upsilon_0 )}.$$
This shows that, if $$|\omega^v_i|>\frac{\|e^{\star}(\widetilde z,\omega^v)\|_{(\Upsilon_0,1)} + \|e^{\star}(\widetilde z,\omega^v)\|_{(\Upsilon_0,2M)}}{\sigma_{min,2M}^2(\Phi^\Xi )},$$ then $\bar \omega\neq 0$ and, consequently
$$
\begin{aligned}
\lambda \doteq &\left\{i: |\omega_i^v|\!>\!\frac{\|e^{\star}(\widetilde z,\omega^v)\|_{(\Upsilon_0,1)}\!+\!\|e^{\star}(\widetilde z,\omega^v)\|_{(\Upsilon_0,2M)}}{\sigma_{min,2M}^2(\Upsilon_0 )}\right\}\\ &\subseteq \text{supp}(\bar \omega).
\end{aligned}$$
If follows that $\|\bar\omega\|_0\geq \text{card}(\lambda),$ which yields \eqref{eqn:thm2.1.ss}.
Now, the constraints \eqref{eqn:cnstr1} and \eqref{eqn:cnstr2} imply that
$$\text{supp}(\widehat\omega)=\{i:|\omega_i^v|\geq \eta(\widehat\omega)\}.$$
Moreover, if condition \eqref{eqn:thm2.ineq.ss} holds, then 
$$
\begin{aligned}
&\{i:|\omega_i^v|\geq \eta(\widehat\omega)\} \\&\subseteq \left\{i: |\omega_i^v| > \frac{\|e^{\star}(\widetilde z,\omega^v)\|_{(\Upsilon_0,1)} + \|e^{\star}(\widetilde z,\omega^v)\|_{(\Upsilon_0,2M)}}{\sigma_{min,2M}^2(\Upsilon_0 )}\right\} \\
&\doteq \lambda \subseteq \text{supp}(\bar \omega)
\end{aligned}
$$
It follows that $\text{supp}(\widehat \omega) \subseteq \text{supp}(\bar \omega)$. But $\bar\omega$ is the sparsest vector satisfying the constraint $\|\Upsilon_0\widetilde z - \Upsilon_0\Phi\omega\|_2\leq \mu$. Thus, relation \eqref{eqn:thm2.res.ss} follows, concluding the proof.

\bibliographystyle{IEEEtran}
\bibliography{main}

\end{document}